\documentclass[useAMS,usenatbib]{modified_mn2e}
\input{psfig}

\def\wp{$w_p(r_p)$}
\def\om{\Omega_m}
\def\spplane{$r_\sigma - r_\pi$}
\def\xx{$\xi(r_\sigma,r_\pi)$}

\def\xg{$\xi_R(r)$}
\def\av{\alpha_v}
\def\avc{\alpha_{vc}}
\def\kmsmpc{km\,s$^{-1}$\,Mpc$^{-1}$}
\def\hmpc{$h^{-1}$Mpc}

\def\etal{{\it et.\,\,al.}}

\def\s8{\sigma_8}
\def\mstar{$M_\ast$}

\def\hmsol{$h^{-1}$M$_\odot$}
\def\kms{km\,s$^{-1}$}
\def\rsig{$r_\sigma$}
\def\rpi{$r_\pi$}
\def\G{\mbox{G}}

\def\rhob{\tilde{\rho}}
\def\rhobtwo{\tilde{\rho}_{200}}
\def\lcdm{$\Lambda$CDM}

\def\gadget{{\scriptsize GADGET}}

\def\mrlogh{M_r-5\log\,h}

\def\asat{{\alpha_{\rm sat}}}
\def\mmin{M_{\rm min}}

\def\mlima{M_{\rm lim,1}}
\def\mlimb{M_{\rm lim,2}}
\def\mcut{M_{\mbox{\scriptsize cut}}}
\def\navg{\langle N\rangle_M}

\def\navga{\langle N\rangle_{M_1}}
\def\navgb{\langle N\rangle_{M_2}}

\def\nsat{\langle N_{\rm sat}\rangle_M}
\def\nsata{\langle N_{\rm sat}\rangle_{M_1}}
\def\nsatb{\langle N_{\rm sat}\rangle_{M_2}}
\def\ncen{\langle N_{\rm cen}\rangle_M}
\def\ncena{\langle N_{\rm cen}\rangle_{M_1}}
\def\ncenb{\langle N_{\rm cen}\rangle_{M_2}}

\def\ngavg{\bar{n}_g}
\def\ngavgp{\bar{n}^\prime_g}

\def\NNm1{\langle N(N-1) \rangle}
\def\NNm1sat{\langle N_{\rm sat}(N_{\rm sat}-1) \rangle}
\def\xis{\xi_{\rm 1h}}
\def\xisz{\xi_{\rm 1h}(r_\sigma,r_\pi)}
\def\xid{\xi_{\rm 2h}^R}
\def\rvir{R_{\rm vir}}
\def\rvira{R_{\rm vir,1}}
\def\rvirb{R_{\rm vir,2}}
\def\rvirc{R_{\rm vir,0}}

\def\intdn{\int_{0}^\infty dM\frac{dn}{dM}}

\def\rhalf{r_{\xi/2}}
\def\xizr{\xi_{0/R}}

\def\xiQp{Q_\xi}

\def\plin{P_{\rm lin}(k)}
\def\Ph{P_{h}}

\def\Pgh{P_{g+h}}

\def\Pm{P_{m}(\delta|r)}

\def\Pmb{P_{m}(\delta|r,M_1,M_2)}
\def\meandelta{\langle \delta\rangle}

\def\rhozero{\tilde{\rho}_0}
\def\rhoone{\tilde{\rho}_1}
\def\rhotwor{\tilde{\rho}_{200,r}}
\def\rhotwot{\tilde{\rho}_{200,r}}

\def\sigsat{\sigma_{\rm sat}}
\def\sigvir{\sigma_{\rm vir}}
\def\sigcs{\sigma_{cs}}

\def\one_col_fig{8.0cm}
\def\two_col_fig{15.0cm}

\title[Redshift-Space Distortions with the HOD]
{Redshift-Space Distortions with the Halo Occupation Distribution  II: Analytic Model}
\author[J.~L.~Tinker]{Jeremy L. Tinker$^{1}$\thanks{E-mail: tinker@cfcp.uchicago.edu}\\ \\
$^{1}$Kavli Insitute for Cosmological Physics, University of Chicago, 933 E. 56$^{th}$ St., Chicago, IL 60637
}

\begin{document}

\date{}

\pubyear{2005}

\maketitle

\begin{abstract}

We present an analytic model for the galaxy two-point correlation
function in redshift space. The cosmological parameters of the model
are the matter density $\om$, power spectrum normalization $\s8$, and
velocity bias of galaxies $\av$, circumventing the linear theory
distortion parameter $\beta$ and eliminating nuisance parameters for
non-linearities. The model is constructed within the framework of the
Halo Occupation Distribution (HOD), which quantifies galaxy bias on
linear and non-linear scales. We model one-halo pairwise velocities by
assuming that satellite galaxy velocities follow a Gaussian
distribution with dispersion proportional to the virial dispersion of
the host halo. Two-halo velocity statistics are a combination of
virial motions and host halo motions. The velocity distribution
function (DF) of halo pairs is a complex function with skewness and
kurtosis that vary substantially with scale. Using a series of
collisionless N-body simulations, we demonstrate that the shape of the
velocity DF is determined primarily by the distribution of local
densities around a halo pair, and at fixed density the velocity DF is
close to Gaussian and nearly independent of halo mass. We calibrate a
model for the conditional probability function of densities around
halo pairs on these simulations. With this model, the full shape of
the halo velocity DF can be accurately calculated as a function of
halo mass, radial separation, angle, and cosmology. The HOD approach
to redshift-space distortions utilizes clustering data from linear to
non-linear scales to break the standard degeneracies inherent in
previous models of redshift-space clustering. The parameters of the
occupation function are well constrained by real-space clustering
alone, separating constraints on bias and cosmology. We demonstrate
the ability of the model to separately constrain $\om$, $\s8$, and
$\av$ in models that are constructed to have the same value of $\beta$
at large scales as well as the same finger of god distortions at small
scales.

\end{abstract}

\begin{keywords}
cosmology: theory --- galaxies: clustering --- large-scale
structure of universe
\end{keywords}


\section{Introduction}

The growth of structure through gravity creates peculiar velocities
with respect to the smooth Hubble flow, making the line of sight a
preferred direction when redshift is used as a measure of
distance. The systematic differences between the distribution of
structure in real and redshift space contain information about both
the cosmology of the universe and the bias of galaxies observed (see,
e.g., \citealt{peebles76, sargent77, kaiser87}). In Paper I
(\citealt{paper1}), we used numerical simulations to develop a
blueprint for determining cosmological parameters from redshift-space
distortions. In this paper, we use those simulations to calibrate an
analytic model for redshift-space galaxy clustering that has the
necessary accuracy to interpret high-precision measurements from
large-scale galaxy redshift surveys, like the Two-Degree Field Galaxy
Redshift Survey (2dFGRS; \citealt{colless01}) and the Sloan Digital
Sky Survey (SDSS; \citealt{york00}).

The effect of redshift-space distortions is most apparent at small
scales, where the internal motions of galaxies in groups and clusters
spread out close pairs along the line of sight, creating the so-called
``fingers-of-god'' (FOGs). This picture of clustering in redshift space
can be expressed as a convolution of the real-space two-point
correlation function \xg\ with the probability distribution function
(PDF) of galaxy pairwise velocities $P(v_z)$, i.e.,

\begin{equation}
\label{e.stream}
1+\xi(r_\sigma,r_\pi) = \int_{-\infty}^\infty [1+\xi_R(r)]\, P(v_z)\, dv_z,
\end{equation}

\noindent where $\xi_R(r)$ is the real-space correlation function,
\rsig\ is the projected separation, \rpi\ is the line-of-sight
separation, $r^2=r_\sigma^2+z^2$, and $v_z = H(r_\pi-z)$, where $H$ is
the Hubble constant. Equation (\ref{e.stream}) is often referred to as
the ``streaming model'' (\citealt{peebles80}). Although this approach
has been used primarily to model observations of small-scale
distortions (e.g., \citealt{peebles79, davis83, bean83}), equation
(\ref{e.stream}) is valid in the linear regime as well
(\citealt{fisher95}). Scoccimarro (\citeyear{roman04}) demonstrated
that, provided $P(v_z)$ is correct, the streaming ``model'' is a valid
description of the relation between the real- and redshift-space
correlation functions on all scales.

As presented, equation (\ref{e.stream}) contains no cosmology. The
aforementioned investigations utilized the streaming model to estimate
the velocity dispersion of galaxy pairs, which was used to estimate
the matter density parameter $\om$ through a ``cosmic virial
theorem''. More recent studies of redshift-space distortions have
utilized a modified linear theory model of anisotropies that, when
applied to the real-space galaxy power spectrum $P_R(k)$, takes the
form

\begin{equation}
\label{e.lin_exp}
P_Z(k,\mu) = P_R(k)(1+\beta\mu^2)^2 (1+k^2\sigma_k^2 \mu^2/2)^{-1},
\end{equation}

\noindent where $\beta=\om^{0.6}/b_g$, $b_g$ is the linear bias
parameter, and $\mu$ is the cosine of the angle between the wavevector
$k$ and the line of sight. The term $(1+\beta\mu^2)^2$, derived from
linear theory by \cite{kaiser87}, models the coherent flow of matter
out of underdense regions and into overdense regions. The last term on
the right hand side represents an exponential distribution of random,
uncorrelated peculiar velocities, which dominates $P_Z$ on small
scales and is meant to encapsulate the FOG effect described
above. Equation (\ref{e.lin_exp}), commonly referred to as the
``dispersion model'', has two free parameters, $\beta$ and the galaxy
velocity dispersion $\sigma_k$. \cite{roman04} points out several
deficiencies in this model, both in the inability of linear theory to
properly describe anisotropies even in the large-scale limit, and in
the oversimplification of using a single parameter $\sigma_k$, which
has no clear physical definition, to model small-scale
velocities. Consequently, equation (\ref{e.lin_exp}) introduces a
$10-15\%$ systematic error in the determination of $\beta$
(\citealt{hc99}; Paper I), a level of error significant compared with
the precision achievable with SDSS and the 2dFGRS.

The goal of this paper is to create an analytic model for the
redshift-space correlation function by combining the streaming model
of equation (\ref{e.stream}) with the Halo Occupation Distribution
(HOD; see, e.g., \citealt{jing_etal98, ma00, peacock00, seljak00,
  benson01, roman01, bw02, cooraysheth02}). The HOD quantifies bias on
both linear and non-linear scales for a given galaxy sample by
specifying the probability $P(N|M)$ that a halo of mass $M$ contains
$N$ galaxies of a given type, together with any spatial and velocity
biases between galaxies and dark matter within individual halos. The
HOD has been utilized to model the real-space clustering of galaxies
in the SDSS (\citealt{zehavi04a, zehavi04b, tinker05a}) and the 2dFGRS
(\citealt{tinker_pvd06}; see also \citealt{yang03}). In this paper we
extend the HOD model from real space to redshift space by
providing a model for $P(v_z)$ which is physically motivated and
empirically calibrated on numerical simulations.

Several recent papers have presented calculations of redshift-space
distortions using halo models of dark matter and galaxy clustering
(\citealt{seljak01, white01, kang02, cooray04, skibba05, slosar06}),
providing insight into the role of non-linear dynamics and non-linear
bias in shaping clustering and anisotropy. However, these studies rely
on the same linear theory component of equation (\ref{e.lin_exp}) for
large-scale anisotropies. \cite{kang02} show that their model
only reproduces the dark matter $P_Z(k,\mu)$ from N-body simulations
after introducing a $\sigma_k$ parameter for the {\it halos}, even
after the virial motions of particles within halos were taken into
account. \cite{skibba05} demonstrate the difficulty is modeling
redshift-space galaxy clustering in the transition region between
quasi-linear and fully non-linear regimes using a linear theory
description of halo velocities. Other recent papers have used the halo
approach to model galaxy and dark matter velocity statistics
(\citealt{sheth01a, sheth01b, sheth01c}). While the model outlined in
these papers is derived from first principles, in contrast to the
calibrated model presented here, it is still based on linear theory,
which does not provide the required accuracy for a robust
implementation of equation (\ref{e.stream}). The purpose of our model
is less as a first-principles derivation of $P(v_z)$ than as a tool to
extract information from forthcoming observational data. In this
context, the accuracy of the model is the paramount concern. In the
course of developing the model, we will also gain new insight into the
physics that determines $P(v_z)$, especially the role of environment
in producing a non-Gaussian velocity distribution.

An accurate model for \xx\ with the HOD must properly incorporate halo
motions. A proper model for halo pairwise velocities $P_h(v)$ must
correctly describe the distribution function for an arbitrary pair of
halo masses, at any angle with respect to the line of sight, and as a
function of separation. In the large-scale limit, linear theory is
adequate for describing the mean infall velocities of halos (see,
e.g., \citealt{juszkiewicz99, sheth01a}). However, the applicability
of linear theory is problematic at scales where the observational data
are robust. At all scales, linear theory does not accurately predict
the pairwise dispersion (\citealt{roman04}). Higher order moments also
play an important role in $\Ph(v)$. N-body results have shown that the
radial velocity PDF of dark matter halos exhibits significant skewness
and kurtosis (\citealt{zurek94, juszkiewicz98}).\footnote{In this
  paper we use the convention of {\it radial} being the direction
  connecting the halo pair, {\it tangential} being in a direction
  orthogonal to the radial, and {\it line-of-sight} to be the
  direction from the observer. Velocities in these directions will be
  referred to as $v_r$, $v_t$, and $v_z$, respectively.} The skewness
arises from the infall of matter into overdense regions
(\citealt{juszkiewicz98}). The kurtosis, manifesting as exponential
wings in both the radial and tangential velocities, is due to local
non-linear effects for each halo in the pair. \cite{roman04} concludes
that a Gaussian is {\it never} a good description of velocities, even
at the largest scales. \cite{roman04} focuses on velocity statistics
of dark matter, but local non-linear effects apply to halos as well
(\citealt{kang02}), and $P_h(v)$ from simulations are non-Gaussian at
all scales. It is not sufficient for $P_h(v)$ to describe the first
two moments of the velocity distribution. To accurately model \xx,
$P_h(v)$ must reasonably describe higher order moments of the
distribution as well (see, e.g., \citealt{fisher94}).

As stated in Paper I, our method for analyzing redshift-space
distortions is to first use measurements of the projected correlation
function \wp\ to determine the parameters of the HOD for a given
cosmology. If HOD parameters cannot be found that allow the
cosmological model to reproduce the observed \wp\ then that model is
ruled out. Once the HOD has been determined, the redshift-space
clustering is investigated by the analytic model presented here or the
N-body approach of Paper I. The cosmological parameters that most
directly influence redshift-space clustering are the matter density
parameter $\om$, the amplitude of the linear matter power spectrum,
defined here by $\s8$, the rms linear-theory mass fluctuation in 8
\hmpc\ spheres (where $h \equiv H_0/100$ \kmsmpc), and the velocity
bias of the galaxy sample, which we parameterize by $\av$, the ratio
between the satellite galaxy velocity dispersion and the virial
velocity dispersion of the dark matter halo. One can consider models
in which the velocities of central galaxies are biased with respect to
mean motion of the host halo. In Paper I we considered the shape of
the linear matter power spectrum $\plin$ to be determined by
measurements of the cosmic microwave background (CMB) anisotropies and
the large-scale galaxy power spectrum (e.g., \citealt{percival02,
  spergel03, tegmark04a}).  An advantage of the analytic model for
\xx\ is that it allows for marginalization over $\plin$ without the
cumbersome process of running multiple N-body simulations. The
redshift-space observables explored here and in Paper I are largely
independent of $\plin$, thus the constraints on HOD parameters come
primarily from \wp\ and to a small degree the assumed $\plin$, and the
constraints on our cosmological parameter space of $\om$, $\s8$, and
$\av$ come from the redshift-space data. Any degeneracies between
cosmology and HOD parameters can be identified and marginalized over
as well.

In \S 2 we outline the analytic model, presenting analytic
expressions for \xx\ by combining the HOD model for real-space
clustering with the streaming model of equation (\ref{e.stream}). In
\S 3 we detail the model for halo pairwise velocities, the key
ingredient in the two-halo term, and calibrate it on the N-body
simulations. \S 4 tests the model against the mock galaxy samples of
Paper I, demonstrating that the model can recover the correct
cosmological parameters from a wide variety of cosmologies which
produce significant degeneracies in redshift-space clustering. In \S 5
we summarize and discuss prospects for applying the model to current
data sets.


\begin{figure}
\centerline{\psfig{figure=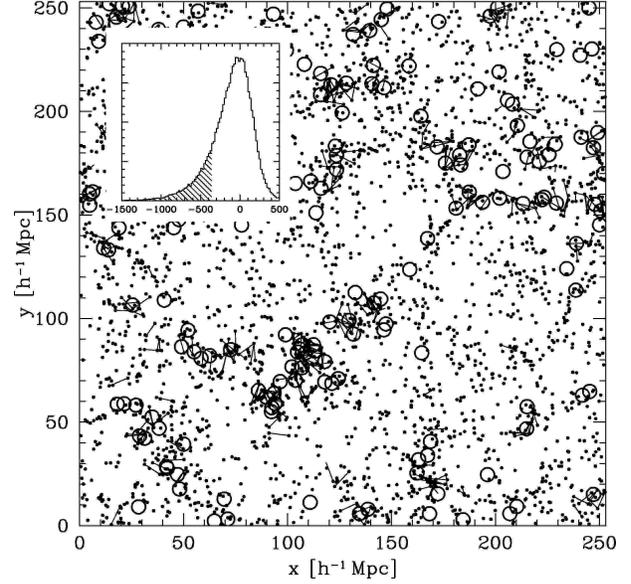,width=\one_col_fig}}
\caption{ \label{slice}
A slice through one simulation showing the locations of halos of
different masses. Large open circles are halos of mass $M > 3\times
10^{13}$. Small filled circles are halos of mass $\sim 2\times 10^{12}$
\hmsol. The inset box plots the probability distribution function (PDF)
of radial velocities for the lower-mass halo pairs in the slice. The
shaded regions of the histogram highlights the skewness toward
high-velocity pairs. All halo pairs within the shaded region of the
histogram are connected in the slice plot.}
\end{figure}


\section{Modeling the correlation function in real and redshift space}

The real-space galaxy correlation function \xg\ can be calculated
analytically with the HOD for a given galaxy sample and cosmology. In
the HOD framework, galaxy clustering is separated into two distinct
parts; pairs of galaxies within a single halo and those from two
separate halos. The total correlation function is the sum of these two
contributions (see, e.g., \citealt{bw02, cooraysheth02, zheng04a}; and
\citealt{tinker05a} for details on calculating the real-space
correlation function).

The real-space one-halo term is written as

\begin{eqnarray}
\label{e.xi1h}
1+\xis(r) &=& \frac{1}{2\pi r^2\ngavg^2}\intdn \frac{\langle N(N-1)\rangle_M}{2} \nonumber \\ & & 
              \times\frac{1}{2\rvir(M)} F^\prime\left(\frac{r}{2\rvir}\right),
\end{eqnarray}

\noindent where $\ngavg$ is the mean number of density of galaxies,
$dn/dM$ is the halo mass function (here we use the formula presented
in \citealt{jenkins01}), and $\langle N(N-1) \rangle_M$ is the average
number of galaxy pairs in a halo of mass $M$. The function
$F^\prime(x)$ is the fraction of galaxy pairs at the scaled radius
$x\equiv r/2\rvir$. In practice, $F^\prime(x)$ is different for galaxy
pairs which involve the central galaxy and those between two satellite
galaxies, and the calculation is separated into these two
terms.


\begin{figure*}
\vspace{-1.5cm}
\centerline{\psfig{figure=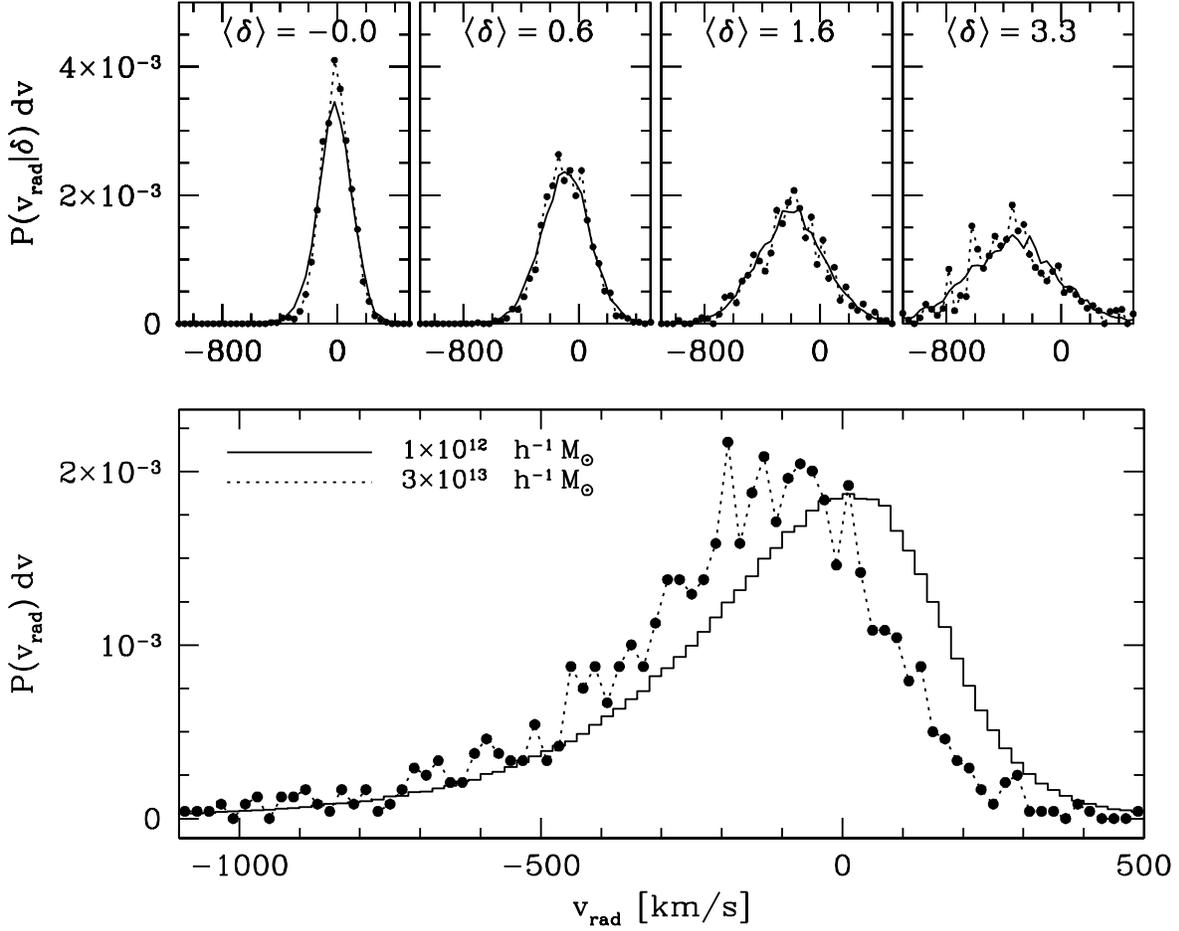,width=17.0cm}}
\vspace{-2cm}
\caption{ \label{vbin_by_delta}
Velocity PDFs for halo pairs at different local densities. The top four
panels plot PDFs as a function of local environment for low-mass halos
(solid histograms) and high-mass halos (filled circles connected with
dotted lines). The local density is calculated by a top-hat smoothing
kernel centered on the midpoint between the halo pair, with radius equal
to the halo separation. The bottom panel shows the unconditional
velocity PDF for the same halos. }
\end{figure*}

In real space, one-halo pairs dominate \xg\ at small scales. In
redshift space, these pairs have large relative motions due to the
virial dispersion of the halos they occupy, spreading these pairs out
along the line of sight. We model the velocity distribution within
each halo as an isotropic, isothermal Gaussian distribution, an
approximation supported by numerical hydrodynamic simulations
(\citealt{faltenbacher04}) and observational analysis of rich SDSS
clusters (McKay et al., in preparation). The satellite galaxy velocity
dispersion in a halo of mass $M$ is proportional to the virial
dispersion,

\begin{equation}
\label{e.sigma_gal}
\sigma_{\rm sat}^2 = \av^2\,\sigma_{\rm vir}^2 = \av^2\, \frac{\G M_h}{2R_{\rm vir}}.
\end{equation}

\noindent
We define $\rvir$ to be the radius at which the mean interior density
of the halo is 200 times the background density, $\rvir =
(4M_h/3\pi\bar{\rho}200)^{1/3}$. The constant of proportionality $\av$
is the galaxy velocity bias. For satellite pairs, the dispersion is
$\sigma_{ss} = \sqrt{2}\sigma_{\rm sat}$. For central galaxies, a
natural assumption is that these galaxies are at rest with respect to
the center of mass of the halo, or $\avc=0$. In Paper I we tested this
assumption, demonstrating that $\avc=0.2$ has a negligible effect on
redshift-space observables. We define $\sigma_{\rm cen} =
\avc\sigma_{\rm vir}$, so the dispersion of central-satellite pairs is
$\sigma_{cs} = (\av^2 + \avc^2)^{1/2}\sigma_{\rm vir}$. Deviations
from isotropy have little effect on \xx\ for any physically reasonable
value of the anisotropy, while non-isothermality resulting from
spatial bias or unrelaxed systems can be modeled as an ``effective''
velocity bias parameter (Paper I).

In equation (\ref{e.xi1h}), the total number of one-halo pairs
involves an integral over the halo mass function. At a given
separation, the velocity PDF for one-halo pairs is a superposition of
Gaussians weighted by the relative number of galaxy pairs from each
halo of mass $M$. This superposition of Gaussians of varying
dispersion makes the overall pairwise galaxy velocity dispersion
approximately exponential (\citealt{sheth96}), in accord with
observational inferences (\citealt{davis83, hawkins03}).

The combination of equations (\ref{e.stream}) and (\ref{e.xi1h}) for
central-satellite pairs is

\begin{eqnarray}
\label{e.xi1h_cs}
\xis^{(cs)}(r_\sigma,r_{\pi}) & = & \frac{1}{2\pi \bar{n}_g^2} 
  \int_{0}^{\infty} dM \frac{dn}{dM} \frac{\nsat\ncen}{2\rvir} \nonumber \\
& & 
  \int_{-\infty}^{\infty} F_{cs}^\prime \left( \frac{\sqrt{r_\sigma^2+z^2}}{2\rvir} \right)
  \frac{1}{\sqrt{2\pi}\sigcs} \nonumber \\ & & \times
  \exp \left[ \frac{-(r_\pi - z)^2}{2\sigcs^2}\right]
  \frac{dz}{(r_\sigma^2 + z^2)},
\end{eqnarray}

\noindent where $\ncen$ and $\nsat$ are the mean number of central and
satellite galaxies for mass $M$, respectively. $F_{cs}^\prime(x)$ is
proportional to the halo density profile, which we assume to follow
the form of Navarro et al.~(1997, hereafter NFW). For
satellite-satellite pairs, the one-halo term is

\begin{eqnarray}
\label{e.xi1h_ss}
\xis^{(ss)}(r_\sigma,r_{\pi}) & = & \frac{1}{2\pi \bar{n}_g^2} 
  \int_{0}^{\infty} dM \frac{dn}{dM} \frac{\NNm1sat}{4\rvir} \nonumber \\
& & 
  \int_{-\infty}^{\infty} F^\prime_{\rm ss} \left( \frac{\sqrt{r_\sigma^2+z^2}}{2\rvir} \right)
  \frac{1}{\sqrt{\pi}2\sigsat} \nonumber \\ & & \times
  \exp \left[ \frac{-(r_\pi - z)^2}{4\sigsat^2}\right]
  \frac{dz}{(r_\sigma^2 + z^2)},
\end{eqnarray}

\noindent 
where $\NNm1sat$ is the second moment of the satellite occupation
function, and $F_{ss}^\prime(x)$ is the fraction of
satellite-satellite pairs (see \citealt{sheth01a} for a derivation of
$F_{cs}(x)$ and $F_{ss}(x)$ for the NFW profile).  The full one-halo
correlation function is $\xisz = \xis^{(cs)} + \xis^{(ss)}$.

The two halo term is most straightforward to calculate through a direct
implementation of the equation (\ref{e.stream}),

\begin{equation}
\label{e.stream2h}
1+\xi_{\rm 2h}(r_\sigma,r_\pi) = \int_{-\infty}^\infty [1+\xid(r)]\, P_{\rm 2h}(v_z|r,\phi)\, dv_z.
\end{equation}

\noindent 
where $\xid(r)$ is the two-halo contribution to the real-space
correlation function, $P_{\rm 2h}(v_z|r,\phi)$ is the line-of-sight
velocity PDF of galaxy pairs from two distinct halos, and $\cos \phi =
r_\sigma/r$. This PDF is a convolution of the line of sight PDF of
halo pairwise velocities with the Gaussian distribution for galaxy
velocities within each halo. $P_{\rm 2h}(v_z|r,\phi)$ is a
pair-weighted average integrated over all possible combinations of
halos;

\begin{eqnarray}
\label{e.vpdf}
\lefteqn{P_{\rm 2h}(v_z|r,\phi) = 2(\ngavgp)^{-2} \int_0^{\mlima} dM_1
\frac{dn}{dM_1} \navga} \nonumber \\ & &
\int_0^{\mlimb} dM_2 \frac{dn}{dM_2} \navgb
P_{\rm g+h}(v_z|r,\phi,M_1,M_2) \,, 
\end{eqnarray}

\noindent 
where $\Pgh$ is the velocity PDF for galaxy pairs from halos of masses
$M_1$ and $M_2$. $\Pgh$ includes both the internal motions of the
galaxies within each halo and the relative motions of the halo centers
of mass. The limits on each integral in equation (\ref{e.vpdf}) are
determined by halo exclusion; halo pairs cannot be closer than the sum
of their virial radii (i.e., $\rvira + \rvirb \le r$). At large
separation, this effect is negligible because the largest collapsed
objects have radii $\sim 2$ \hmpc, and the limits of equation
(\ref{e.vpdf}) approach infinity when $r$ is large. At small scales,
halo exclusion strongly influences the number of pairs. In equation
(\ref{e.vpdf}), $\mlima$ is the maximum halo mass such that
$\rvir(\mlima) = r-\rvir(\mmin)$ and $\mlimb$ is related to $M_1$ by
$\rvir(\mlimb) = r-\rvir(M_1)$, where $\mmin$ is the minimum mass halo
that can host a galaxy. The restricted number density, $\ngavgp$, is
the total number of two-halo galaxy pairs at separation $r$, also
calculated using halo exclusion by

\begin{eqnarray}
\label{e.ng_sph}
\ngavgp{^2} =  \int_{0}^{\mlima} dM_1\frac{dn}{dM_1}
   \navga \int_{0}^{\mlimb} dM_2\frac{dn}{dM_2} \navgb.
\end{eqnarray}

\noindent
We denote the velocity PDF of halo pairs as
$\Ph(v_z|r,\phi,M_1,M_2)$. This represents the center-of-mass motions
only. Once the relative motions of the dark matter halos are
determined, $\Pgh$ is calculated by convolving $\Ph$ with the Gaussian
describing the internal motions of the galaxies within the halos. The
model for $\Ph(v_z|r,\phi,M_1,M_2)$ is important both for capturing
the behavior of the redshift-space correlation function in the
transition region between the one- and two-halo terms and for
correctly modeling distortions between quasi-linear to linear
scales. We describe it in detail in the next section.


\begin{figure*}
\centerline{\psfig{figure=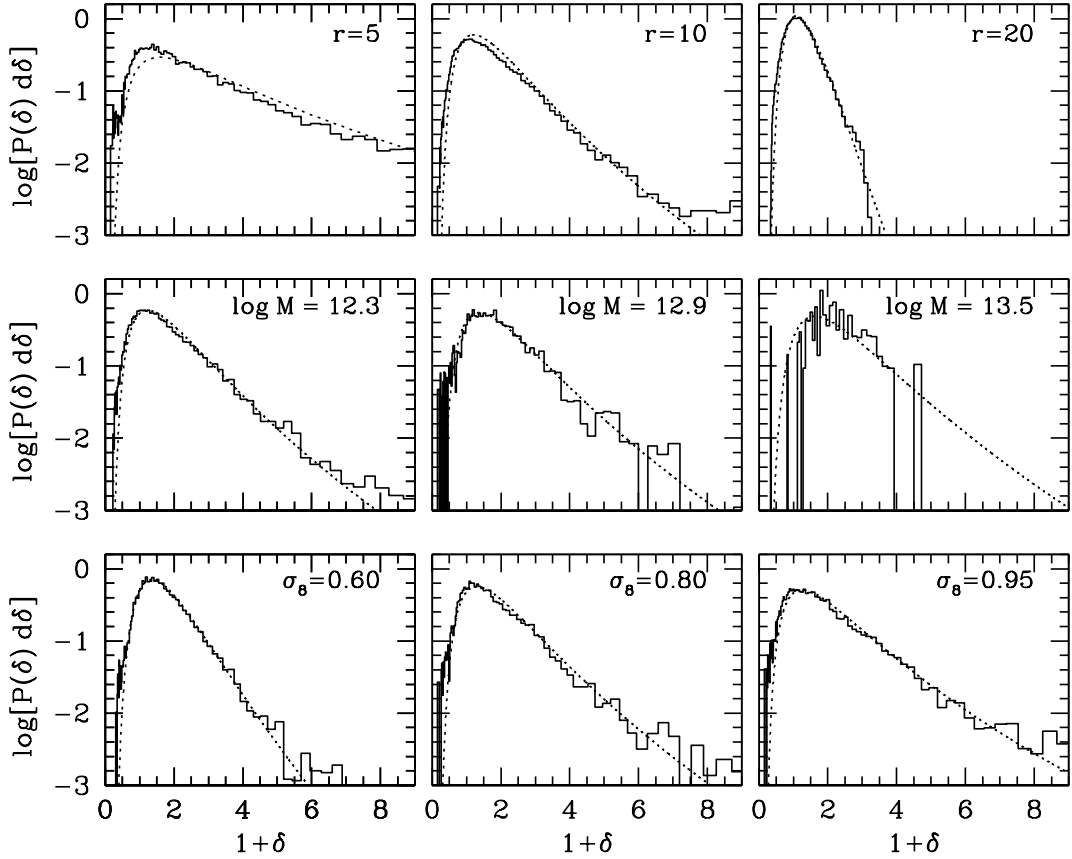,width=\two_col_fig}}
\vspace{-1.5cm}
\caption{ \label{matter_pdf}
Comparison between the conditional matter PDFs measured from the
simulations ({\it solid histograms}) and the model of Equation
(\ref{e.density_pdf}) ({\it dotted lines}). Top row: $\Pmb$ for fixed
mass ($10^{12}$ \hmsol) and $\s8=0.8$, but changing radius. Middle row:
$\Pmb$ for fixed radius (10 \hmpc) and $\s8=0.8$, but increasing halo
mass. Bottom row: $\Pmb$ for fixed mass ($10^{12.3}$
\hmsol) and radius (10 \hmpc), but changing $\s8$.  }
\end{figure*}


\section{The halo velocity model}

\subsection{Conditional matter PDF}

In \S 1 we listed the requirements an accurate model of $P_h(v)$ must
satisfy. To meet these requirements, we use the N-body simulations
from Paper I to explore halo velocity statistics. Our simulation set
consists of five realizations of an inflationary \lcdm\ cosmology with
$\om=0.1$ and $\s8=0.95$ at $z=0$, using the public N-body code
\gadget\ (\citealt{gadget}). We utilize outputs at higher redshifts to
represent models at different values of $\s8$ and $\om$. Each
simulation is $360^3$ particles in a cubic volume $253$ \hmpc\ on a
side, yielding a mass resolution of $9.64\time 10^{10}\times \om$
\hmsol. The simulations have a power spectrum shape parameter
$\Gamma=0.2$ in the parameterization of \cite{ebw92}, and a spectral
index $n_s=1$. See Paper I for further details regarding the
simulations. Our fiducial cosmology is $\om=0.3$, $\s8=0.8$,
corresponding to $z=0.55$. Unless otherwise stated, N-body results
will be using this output.

Figure \ref{slice} elucidates the complexity of
$P_h(v|r,\phi,M_1,M_2)$. The inset box shows the PDF of radial
velocities of halo pairs of mass $M_1=M_2 \approx 2\times 10^{12}$
\hmsol\ at a separation of $r\sim 10$ \hmpc. Negative radial
velocities are defined as the halos going toward one another. The PDF
is not well described by either a Gaussian or an exponential. The
shaded regions of the PDF highlights the skewed tail of halo pairs
rapidly approaching each other. The outer panel plots the positions of
these halos in a 40 \hmpc\ slice through one realization. Small,
filled circles indicate the $2\times 10^{12}$ \hmsol\ halos of the
inset histogram. The open circles in the plot represent high mass
halos, $M>3\times 10^{13}$ \hmsol.  All pairs inside the shaded region
of the histogram have been connected with a line.  These high-velocity
pairs are almost exclusively in regions that include a high mass
halo. Because high mass halos preferentially occupy dense regions, we
recover the well-known result that the non-linear halo velocity field
is coupled to the non-linear density field. However, Figure
\ref{slice} leads us to investigate the velocity PDF as a function of
environment.  In Figure \ref{vbin_by_delta}, the lower panel plots the
same PDF from Figure \ref{slice} for $10^{12}$ \hmsol\ halos at $r\sim
10$ \hmpc. This panel also plots the PDF for $3\times 10^{13}$
\hmsol\ at the same separation. The width of these mass bins is a
factor of two. The PDF for high-mass halos is noisier due to lower
statistics, but it is apparent that the PDF for high-mass halos peaks
at $v \sim -150$ \kms, while the mode of the low-mass PDF is at
zero. It is unlikely to find high-mass halos moving away from one
another at this scale, while that probability is nearly 40\% for
low-mass halos.


\begin{figure}
\centerline{\psfig{figure=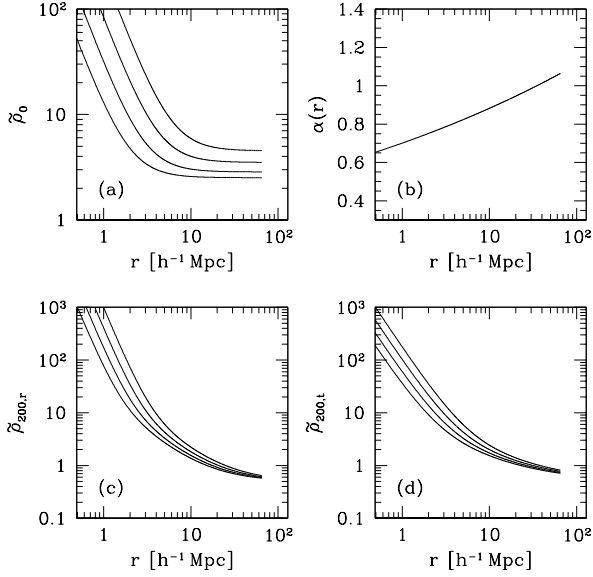,width=\one_col_fig}}
\caption{ \label{params}
The scale and mass dependence of parameters of the two-halo velocity
model. Each line represents of halo pair with mass ratio $M_2/M_1=4^i$,
with $i=0-3$ and $M_1=10^{12}$ \hmsol. Panel (a): cutoff density scale
for the 2-halo conditional matter PDF (Eq.~[\ref{e.rho0}]). Panel (b):
Power-law index for the density-dependence of the pairwise velocity
dispersion (Eq.~[\ref{e.alpha}]). There is only one line in this panel
because there is no dependence on mass. Panel (c): Normalization of the
density dependence of the pairwise radial velocity dispersion
(Eq.~[\ref{e.rho200r}]). Panel (d): Normalization of the density
dependence of the pairwise tangential velocity dispersion
(Eq.~[\ref{e.rho200t}]).}
\end{figure}

The upper panels of Figure \ref{vbin_by_delta} show the PDFs for the
same halos, now binned by local overdensity, $\delta
\equiv\rho/\bar{\rho} - 1$. We calculate $\delta$ inside spheres of
radius equal to the separation of the halo pairs, centered on the
midpoint between the two halos. Setting the radius of the smoothing
kernel equal to the halo separation allows for cleaner determination
of the pair's environment. The four top panels span a range in mean
local density of $\langle \delta\rangle \sim 0$ to 3. From these data
the origin of the high-velocity tail is clear; the negative skewness
arises from halo pairs which lie in dense environments. At $\delta=0$,
the velocity distribution is narrow and peaked at $v\approx 0$
\kms. As $\delta$ increases, both the mean velocity and the dispersion
increase as well. At fixed $\delta$, the PDFs no longer show strong
skewness or kurtosis; a Gaussian is a good approximation.  Therefore
the total distribution in the bottom panel can be modeled as a
superposition of Gaussians weighted by the probability of finding a
halo pair at a given $\delta$.

At a given density, Figure \ref{vbin_by_delta} also demonstrates that
the PDFs for high- and low-mass halos are very similar. It is the
different distributions of local environments which leads to the
differences in the total PDFs in the lower panel. We therefore propose
the ansatz

\begin{eqnarray}
\label{e.ansatz}
\lefteqn{\Ph(v_{[r,t]}|r,M_1,M_2) =}\nonumber \\ & & 
 \int \,P_G(v_{[r,t]}|r,\delta,M_1,M_2)\,P_{m}(\delta|r,M_1,M_2) \,d\delta,
\end{eqnarray}

\noindent 
to model the radial and tangential pairwise motions of halos.  In
equation (\ref{e.ansatz}) $P_G$ is a Gaussian, and
$P_{m}(\delta|r,M_1,M_2)$ is the conditional matter density PDF in
spheres of radius $r$ centered on a halo pair of masses $M_1$ and
$M_2$. An ansatz of this type has been used to model one-point
velocity statistics of halos and galaxies (\citealt{sheth01b,
  hamana03}). Although the PDFs in the upper panels of Figure
\ref{vbin_by_delta} appear independent of halo mass, $P_G$ varies
moderately with mass at small scales because at small $r$ the halos
themselves contribute a large fraction of the total local mass. At
scales significantly larger than the halo radii, $P_G$ becomes
independent of mass.

The PDF for the unconditional matter distribution $\Pm$ is well fit by
a lognormal distribution function with dispersion that varies with
smoothing scale (e.g., \citealt{coles91, kofman94}). The relationship
between the halo mass function and the large-scale density field can
also be quantified (e.g., \citealt{bond91, sheth02, pavlidou05}). The
conditional mass function can be roughly approximated as $n(M|\delta)
\approx [1+b(M)\,\delta]\,n(M)$ (\citealt{mowhite96}), where $b(M)$ is
the large-scale bias factor for halos of mass $M$ (here we use the
halo bias formula in Appendix A of \citealt{tinker05a}). The
distribution of densities at which a halo of a given mass can be found
is roughly $P_{m}(\delta|r,M) \approx [1+b(M)\,\delta]\,P_m(\delta|r)$
(\citealt{hamana03}). The term $[1+b(M)\,\delta]$ effectively shifts
the matter PDF to higher densities by an amount which depends on the
bias of the halo. 

The density distribution around pairs of halos, denoted as $\Pmb$, is
a murky theoretical problem with no obvious straightforward
approach. To model $\Pmb$ we truncate the unconditional matter
distribution with an exponential cutoff at low densities, i.e.,

\begin{equation}
\label{e.density_pdf}
P_{m}(\delta|r,M_1,M_2) = A\, \exp\left[ -\frac{\rhozero(r,M_1,M_2)}{\rhob}\right]\,P_m(\delta|r),
\end{equation}

\noindent 
where $A$ is a normalization constant to make the total probability
unity, $\rhob\equiv 1+\delta$, and $\rhozero$ is a density cutoff
scale to be calibrated from the simulations. The one-point non-linear
distribution function for the dark matter takes the form

\begin{equation}
\label{e.lognormal}
P_m(\delta|r) = \frac{1}{2\pi\sigma_1^2} \, \exp \left[- 
\frac{[\ln(1+\delta) + \sigma_1^2/2]^2}{2\sigma_1^2} \right] \frac{1}{1+\delta},
\end{equation}

\noindent where $\sigma_1^2(r) = \ln [1+\sigma_m^2(r)]$ and $\sigma_m(r)$
is the mass variance in top-hat spheres of radius $r$ (e.g.,
\citealt{coles91, kofman94}). To calculate $\sigma_m(r)$, we use the
non-linear matter power spectrum of \cite{smith03}. The cutoff density
scale $\rhozero$ is

\begin{equation}
\label{e.rho0}
\rhozero = \rhoone\,\left[b(M_1) + b(M_2)\right] + \left(\frac{r}{r_0}\right)^{\alpha_0}.
\end{equation}

\noindent At large scales, equation (\ref{e.rho0}) is proportional to
the sum of the halo bias factors. At small scales, where the masses of
the halos themselves contribute significantly to local density, the
mass contained within $r$ cannot be below $M_1+M_2$. Thus the cutoff
density should scale roughly as $\rhozero \propto
(r/\rvir)^{-3}$. From our N-body simulations, we find $\rhoone=1.41$,
$\alpha_0=-2.2$, and $r_0=9.4 \rvira$, where $\rvira$ is the virial
radius of the larger halo of the halo pair, provide a good fit to the
data. Figure \ref{matter_pdf} compares equation (\ref{e.density_pdf})
to $\Pmb$ measured from the N-body simulations. The top row of panels
plots $\log[P(\delta)]$ measured in spheres centered on halo pairs of
separations 5, 10, and 20 \hmpc, for halos of mass $10^{12}$
\hmsol. The N-body results, shown with the solid histograms, are
lognormal in shape like the unconditional $P_m(\delta)$, with a
dispersion that narrows as the smoothing scale increases. Dotted lines
show equation (\ref{e.density_pdf}), with $\rhozero$ calculated from
equation (\ref{e.rho0}) using the halo bias function of
\cite{tinker05a}.  Equation (\ref{e.density_pdf}) accurately tracks
the change in $\Pmb$ with smoothing scale.

The middle row of panels in Figure \ref{matter_pdf} compares the model
to the N-body results for different halo masses at the same smoothing
scale of 10 \hmpc. N-body halo pairs of mass $\sim 2\times 10^{12}$
\hmsol, which have large-scale bias of $\sim 0.8$ in our fiducial
cosmology, reside on average in overdense regions, with $\meandelta =
0.80$. As the mass increases by factors of 4 and 16 (with $b\sim$ 1.1
and 1.5, respectively), $\meandelta$ increases to 1.00 and 1.36. The
mean density calculated from equation (\ref{e.density_pdf}) for each
halo mass is 0.82, 1.02, and 1.32, in good agreement with the
numerical results.


\begin{figure}
\centerline{\psfig{figure=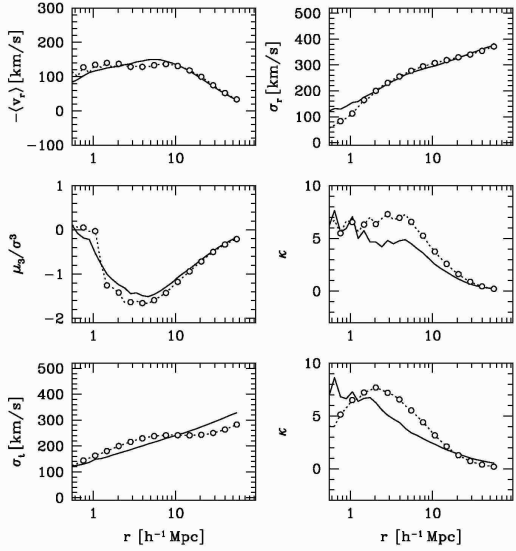,width=\one_col_fig}}
\caption{ \label{pv_moments} Comparison between the velocity
  statistics of N-body halos and the velocity model integrated over
  all halo masses. In all panels the solid line plots the N-body
  results and the open circles connected by dotted lines plots the
  model. The top four panels present the first four moments of the
  radial velocity PDFs as a function of scale: mean $\mu$, dispersion
  $\sigma_r$, skewness $\mu_3/\sigma^3$, and kurtosis $\kappa$. The
  bottom two panels show the dispersion $\sigma_t$ and kurtosis
  $\kappa$ of the tangential PDFs. For proper comparison, the
  integrals over velocity for the moments are binned in the same
  manner as the N-body results and are truncated at the high and low
  velocities of the N-body statistics. }
\end{figure}

The bottom panel in Figure \ref{matter_pdf} isolates the effect of
changing $\s8$ on $\Pmb$. The three panels present numerical results
for $M \sim 5 \times 10^{12}$ \hmsol\ halos at $r \sim 10$ \hmpc\ for
$\s8=0.6$, 0.8, and 0.95. In equation (\ref{e.lognormal}), the
dependence of the matter distribution on the power spectrum is through
the parameter $\sigma_m(r)$, which is proportional to $\s8$. As $\s8$
increases, the dispersion in the unconditional $\Pm$ from equation
(\ref{e.lognormal}) increases. The same is true for $\Pmb$; the N-body
results broaden substantially from $\s8=0.6$ to 0.95. The results from
equation (\ref{e.density_pdf}), plotted once again with the dotted
lines, model the changes of $\Pmb$ with $\s8$ accurately. With three
adjustable parameters, the quantities $\rhoone$, $\alpha_0$ and $r_0$
of equation (\ref{e.rho0}), our model for the conditional matter PDF
$P_{m}(\delta|r,M_1,M_2)$ reproduces the distinct effects of varying
halo mass, smoothing scale, and power spectrum normalization.

\subsection{Parameters of the Gaussian}

For a halo pair at separation $r$ and density $\delta$, we
approximate the radial velocity PDF as a distribution of the
form

\begin{equation}
\label{e.gaussian}
P(v_r|\delta,r,M_1,M_2) = \frac{1}{\sqrt{2\pi}\sigma_r} 
\exp\left[ \frac{-(v-\mu_r)^2}{2\sigma_r^2} \right].
\end{equation}

\noindent Both the mean $\mu_r$ and the dispersion $\sigma_r$ of the
radial velocity PDF depend on $\delta$. The PDF for tangential
velocities is also a Gaussian, but with dispersion $\sigma_t$ and zero
mean. At large scales, or at negative density contrast, $\mu_r$ is well
described by linear theory,

\begin{equation}
\label{e.mu_linear}
\mu_{\rm lin}(\delta,r) = -H\,r\,\om^{0.6}\,\frac{\delta}{3}.
\end{equation}

\noindent At small scales, $\mu_r$ is approximated by the non-linear
spherical collapse model (see, e.g., \citealt{peacocktext}). In this
model, the dependence of the velocity perturbation on density is
expressed parametrically as

\begin{equation}
\label{e.delta_sc}
\delta = \frac{9\,(\gamma - \sin\gamma)^2}{2\,(1-\cos\gamma)^3}-1
\end{equation}

\begin{equation}
\label{e.velocity_sc}
u = \frac{3\sin\gamma\,(\gamma-\sin\gamma)}{(1-\cos\gamma)^2}-1.
\end{equation}

\noindent The mean radial velocity is

\begin{equation}
\label{e.mu_sc}
\mu_{\rm sc}(\delta,r) = H\,r\,\om^{0.6}\,u(\delta)\,\exp[-(4.5/r\rhob)^2],
\end{equation}

\noindent where the exponential term is added to better match the
N-body results at the smallest scales but has little influence on the
overall behavior of $\mu_{\rm sc}$. Investigation of our N-body results
shows that the spherical collapse model best describes the N-body
simulations at $r \la 4$ \hmpc, while linear theory is an excellent
description of the results for $r \ga 20$ \hmpc. In the transition
region, we express $\mu_r$ as a weighted mean of equations
(\ref{e.mu_linear}) and (\ref{e.mu_sc}), which smoothly transitions
between the two regimes. The weighting factor, $w_\mu$ increases
linearly from 0 to 1 in $\ln (r)$ in the range $4 \le r \le 20$ \hmpc,
with $\mu_r = w\mu_{\rm sc} + (1-w)\mu_{\rm lin}$.

Recalling Figure \ref{vbin_by_delta}, the skewness of the radial
velocity PDF arises from the combination of halo pairs in high-density
regions, which have a large streaming velocity and high dispersion,
with halo pairs in mean- to low-density regions, which have low mean
streaming velocity (relative to Hubble flow) and small dispersion.  In
our N-body simulations, the skewness of the radial velocity
distribution does not monotonically increase as halo separation
decreases. As halo pairs become close, of order twice the sum of their
virial radii, the skewness decreases and the PDF becomes more
symmetric. Because $\mu_{\rm sc}$ diverges rapidly as $\delta$ becomes
large, the skewness of the model will always increase with decreasing
$r$. To compensate for this effect, we enforce the condition that at
$r=4\rvira$, $\mu_r$ is held constant for all $\delta$ at the value of
the most probable $\delta$ for that separation, removing the skewness
entirely. In the full halo+galaxy PDF, this change from skewed to
symmetric distribution functions is smoothed out in scale through the
change in $\rvira$ with halo mass.


\begin{figure*}
\centerline{\psfig{figure=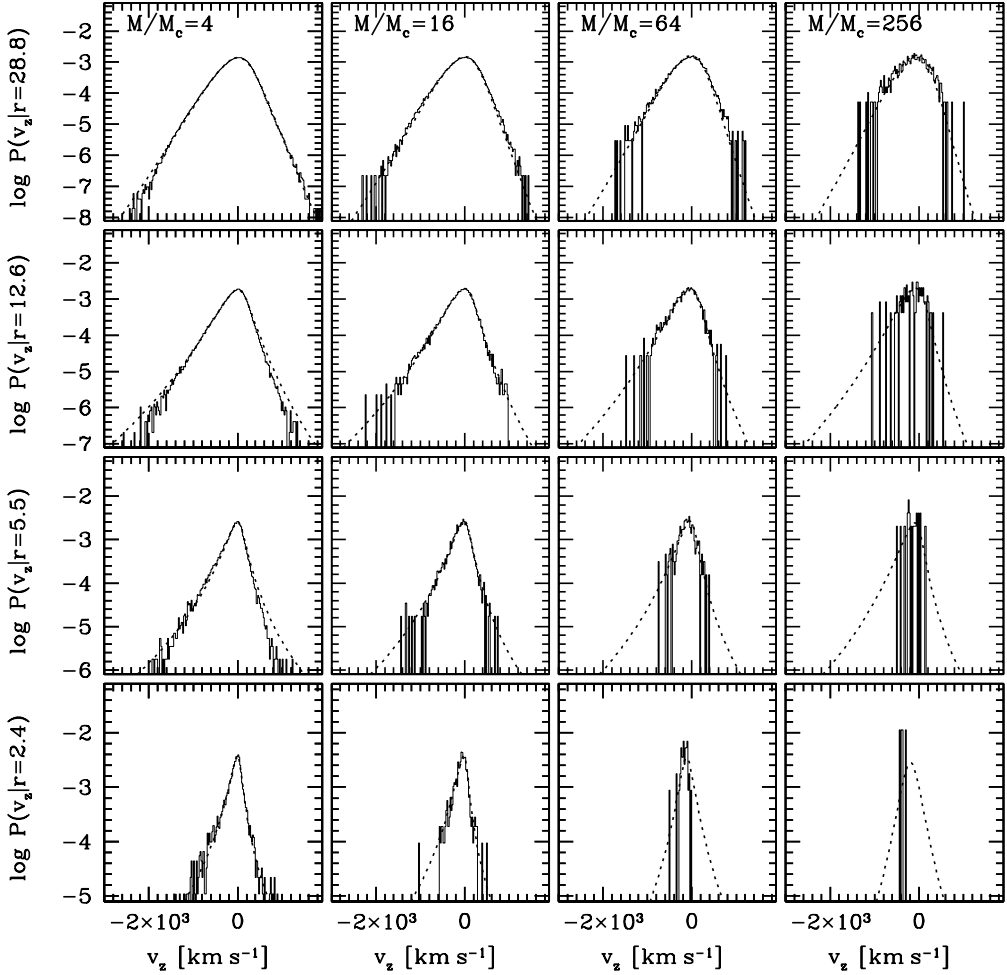,width=15.0cm}}
\caption{ \label{example_pdfs} Comparison between line-of-sight
  velocity PDFs measured from the simulations and from the analytic
  model. Solid histograms show the N-body results and dotted lines
  plot the model calculations. The sixteen panels create a grid in
  which halo mass increases from left to right and separation
  increases from bottom to top. Halo pairs are all of equal masses and
  at an angle of $45^\circ$ with respect to the line of sight. }
\end{figure*}

It is clear from Figure \ref{vbin_by_delta} that the velocity dispersion
also depends on the local density. \cite{hamana03} show that the
one-point velocity dispersion of halos is described by a power law,
i.e. $\sigma_v \propto (1+\delta)^\alpha$. We find that this is also
true of the pairwise velocities but, that both the normalization and
index of the power law are functions of scale. The power-law index
varies weakly with radius, and is modeled by

\begin{equation}
\label{e.alpha}
\alpha(r) = \left(\frac{r}{35\, h^{-1}\,{\rm Mpc}}\right)^{0.1}.
\end{equation}

In linear theory the dispersion of halo space velocities scales with cosmology as
$\om^{0.6}\s8$. We parameterize the halo velocity dispersions in terms
of the overdensity $\rhobtwo$ at which the dispersion is $200$ \kms\ in
our fiducial cosmology:

\begin{equation}
\label{e.sigma_v}
\sigma_{(r,t)} = 200 \left(\frac{\om}{0.3}\right)^{0.6}\left(\frac{\s8}{0.8}\right)
\left(\frac{\rhob}{\rhob_{200\,(r,t)}}\right)^\alpha {\rm km\,s^{-1}},
\end{equation}

\noindent where $\rhob_{200\,(r,t)}$ is a function of $M_1$,
$M_2$, and $r$. We find that the following formulas provide a reasonable
description of the radial and tangential velocity dispersions in the
simulations:

\begin{eqnarray}
\label{e.rho200r}
\lefteqn{\rhob_{200,r}(M_1,M_2,r) =}\nonumber \\ & &  
\left(\frac{r}{5.0\rvira^{1/2}}\right)^{-4.0} + 
\left(\frac{r}{11.5\rvirc^{1/2}}\right)^{-1.3} + \,\,0.50,
\end{eqnarray}

\begin{eqnarray}
\label{e.rho200t}
\lefteqn{\rhob_{200,t}(M_1,M_2,r) =}\nonumber \\ & & 
\left(\frac{r}{7.2\rvira^{1/2}}\right)^{-2.5} + 
\left(\frac{r}{12.6\rvirc^{1/2}}\right)^{-0.8} + \,\,0.48,
\end{eqnarray}

\noindent where $\rvirc = \rvira + \rvirb$. The units of $r$ and
$\rvir$ in equations (\ref{e.rho200r}) and (\ref{e.rho200t}) are
\hmpc. The mass dependence again enters with the virial radii of the
halos, but the dependence on halo mass is weaker than for the density
distribution in equation (\ref{e.rho0}). Figure \ref{params} shows the
radial dependence of the four model parameters which must be
calibrated to the simulations, $\rhozero$, $\alpha$, $\rhotwor$, and
$\rhotwot$, from equations (\ref{e.rho0}), (\ref{e.alpha}),
(\ref{e.rho200r}), and (\ref{e.rho200t}), respectively. The different
lines in each panel are for halo pairs $M_1$ and $M_2$, where
$M_1=10^{12}$ \hmsol, and $M_2/M_1=4^i$, and $i=0-3$. At large scales,
the dependence on halo mass becomes negligible except when it enters
in the bias factor used in the cutoff of the density PDF. At these
scales, the value of $\rhob_{200}$ approaches a constant for both
tangential and radial velocities indicating that the velocity
dispersion becomes independent of scale and mass.


\begin{figure}
\centerline{\psfig{figure=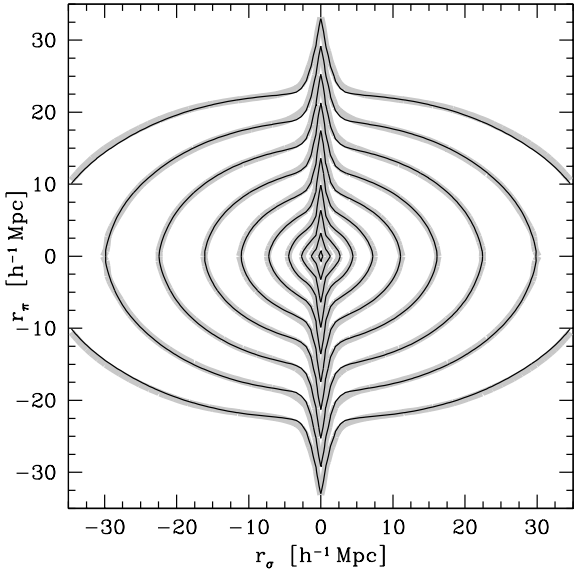,width=\one_col_fig}}
\caption{ \label{xi2d_compare}
\xx\ for HOD3 in Table 1. The contours plot constant amplitude of \xx,
separated by powers of 1.7. The outermost contour is
$1.7^{-5}=0.07$. Thick gray lines show the N-body results and thin black
lines are the model calculation.  }
\end{figure}

Using this model, we integrate over all halo pairs and calculate
equation (\ref{e.vpdf}) for halos alone, without any contribution from
galaxies (i.e.~$\navg=1$). For this special case, equation
(\ref{e.vpdf}) becomes

\begin{eqnarray}
\label{e.halos_only}
P_h(v|r) &=& 2(\bar{n}_h^\prime)^{-2} \int_0^{\mlima} dM_1
\frac{dn}{dM_1} \nonumber \\ & & \int_0^{\mlimb} dM_2 \frac{dn}{dM_2} 
P_{h}(v|r,M_1,M_2)\,
\end{eqnarray}

\noindent where $v$ can refer to either $v_r$ or $v_t$, and
$\bar{n}_h^\prime$ is the number density of halos allowed by halo
exclusion, in analogy to equation (\ref{e.ng_sph}). Equation
(\ref{e.halos_only}) allows a direct comparison to our N-body
simulations. Figure \ref{pv_moments} shows the first four moments of
the halo velocity PDF as a function of scale for the radial
velocities, and the first two even moments for the tangential
velocities, for our fiducial cosmology. The N-body results are shown
with the solid line and the model is the open circles connected by the
dotted line. For calculating the velocity moments with the model, we
use the model value at the center of each bin that has a non-zero
probability in the N-body distribution and set the more extended wings
of $P(v)$ to zero. The contribution from the high-$v$ wings makes
little difference for the comparison of the first two moments, but it
noticeably affects the comparison of the skewness and kurtosis to the
N-body results.

The mean and dispersion of the radial velocities are accurately
described by the model.  For the skewness, denoted as $\mu_3/\sigma^3$
with $\mu_3 = \sum (v_i-\mu)^3/N$, the N-body results demonstrate the
previous statement that the asymmetry is minimal at large separation,
reaches a peak at intermediate scales near 4 \hmpc, and turns over to
become nearly zero at $r\la 1$ \hmpc. The model describes this
behavior well, reaching a peak at the same location and reducing to
zero at $r=1$ \hmpc. The kurtosis, defined as $\kappa =
\mu_4/\sigma^4-3$ with $\mu_4 = \sum (v_i-\mu)^4/N$, approaches zero
at large scales for both the tangential and radial velocities, and
monotonically increases with decreasing $r$ to a value of $\kappa
\approx 7$ at $r \approx 1$ \hmpc\ (for reference, $\kappa = 2$ for an
exponential distribution). The kurtosis is moderately over-predicted
at intermediate scales in the model, but both the sign and slope of
the trend with radius are correct. As we will show later in \S 5, this
small discrepancy with the fourth moment does not adversely affect the
accuracy of the \xx\ model.

\subsection{$P_h(v_z)$ as a function of angle}

For a halo pair at angle $\phi$, the line of sight relative velocity
is $v_z = v_r \sin \phi + v_t \cos \phi$, where $\cos\phi =
r_\sigma/r$.  Unfortunately, knowing the radial and tangential
velocity PDFs independently does not allow one to compute $P(v_z)$
because the radial and tangential velocities are correlated; if the
radial infall velocity is high, the tangential velocity is also likely
to be high. However, this correlation is accounted for by the density
dependence of the PDFs in our model, and at fixed $\delta$ the
distributions $P(v_r)$ and $P(v_t)$ {\it are} independent. The
line-of-sight distributions can then be computed from the appropriate
convolution,

\begin{eqnarray}
\label{e.rt_convolution}
\lefteqn{P_h(v_z|\phi, \delta, M_1, M_2, r) =} \nonumber \\ & & 
 \int_{-\infty}^{\infty} P_h(v_t|\phi, \delta, M_1, M_2, r)\,
P_h(v_r|\phi, \delta, M_1, M_2, r)dv_t, \nonumber \\
\end{eqnarray}

\noindent where $v_r = (v_z - v_t\,\cos\phi)/\sin\phi$, to
obtain

\begin{eqnarray}
\label{e.pv_angle}
P_h(v_z| \phi, \delta, M_1, M_2, r) &=& 
\frac{1}{\sqrt{2\pi}}\, \left(\sigma_t^2\cos^2\phi + \sigma_r^2\sin^2\phi\right)^{-1/2}\nonumber \\ & & 
\times \exp\left[ \frac{-(v_z-\sin\phi\,\mu_r)^2}
{2(\sigma_t^2\cos^2\phi + \sigma_r^2\sin^2\phi)}\right]. \nonumber \\ & &
\end{eqnarray}

\noindent 
Inspection of equation (\ref{e.pv_angle}) produces the expected result
that at $\phi = \tan\phi = \sin\phi = 0$, the expression reduces to a
zero-mean Gaussian with dispersion $\sigma_t$. For $\phi=\pi/2$, it
reduces to equation (\ref{e.gaussian}).

Rather than compare the moments of $P_h(v_z)$ from the model to the N-body
simulations at a given $\phi$, since they appear much as they do in
Figure \ref{pv_moments}, instead we compare the PDFs themselves for
several masses and separations at an angle of $45^\circ$. The sixteen
panels of Figure \ref{example_pdfs} represent a grid of models where
each row is a different separation and each column is a different
mass. In each column, the halo pairs are $M_1=M_2=4^i\,M_c$, with
$i=1-4$, and $M_c = 10^{12}$ \hmsol. In each panel the solid
histogram represents the N-body results and the dotted line plots the
model calculation. The agreement between the model and the N-body
results is excellent wherever the N-body statistics are good. The model
reproduces the numerical results over an order of magnitude variation in
radius and almost two decades in halo mass.

\subsection{Combining halos and galaxies}

Because the halo PDF at fixed $\delta$, $r$, and halo masses is a
Gaussian, internal motions of the galaxies can be included by adding
the galaxy dispersion to the halo velocity dispersion in
quadrature. As in the one-halo term, pairs involving central and
satellite galaxies must be treated separately once again. In the
two-halo term, there are four distinct types of pairs with four
velocity dispersions: central-central pairs, satellite-satellite
pairs, and satellite galaxies from $M_1$ paired with the central
galaxy of $M_2$ and vice versa. The four velocity dispersions are

\begin{equation}
\label{e.sigma_g+h}
\begin{array}{lll}
\sigma_1^2 & = & \sigma_h^2 + [\sigvir^2(M_1) + \sigvir^2(M_2)]\,\avc^2 \nonumber \\
\sigma_2^2 & = & \sigma_h^2 + [\sigvir^2(M_1) + \sigvir^2(M_2)]\,\av^2 \nonumber \\
\sigma_3^2 & = & \sigma_h^2 + [\avc^2\sigvir^2(M_1) + \av^2\sigvir^2(M_2)] \nonumber \\
\sigma_4^2 & = & \sigma_h^2 + [\avc^2\sigvir^2(M_2) + \av^2\sigvir^2(M_1)] \nonumber, \\
\end{array}
\end{equation}

\noindent 
where $\sigma_h^2 = \sigma_t^2\cos^2\phi + \sigma_r^2\sin^2\phi$. The
complete PDF is a linear combination of these four Gaussian functions,

\begin{eqnarray}
\label{e.pdf_sum}
\lefteqn{P_{g+h}(v_z|\phi,r,M_1,M_2,\delta) =}\nonumber \\ & &  \sum_{i=1}^4\, \
\frac{w_i}{\sqrt{2\pi}\sigma_i\,}\exp\left[ \frac{-(v_z - \mu_r\sin\phi)^2}{2\sigma_i^2}
\right],
\end{eqnarray}

\noindent 
where each term is weighted by the relative number of pairs of each
type, i.e.,

\begin{equation}
\label{e.w_i}
\begin{array}{lll}
w_1 & = & \ncena\,\ncenb/\navga\,\navgb \\
w_2 & = & \nsata\,\nsatb/\navga\,\navgb \\
w_3 & = & \ncena\,\nsatb/\navga\,\navgb \\
w_4 & = & \nsata\,\ncenb/\navga\,\navgb . \\
\end{array}
\end{equation}

To obtain the full galaxy+halo PDF for a pair of halos $M_1$ and
$M_2$, equation (\ref{e.pdf_sum}) is weighted by
$P_m(\delta|r,M_1,M_2)$ and integrated over all densities, i.e.,

\begin{eqnarray}
\label{e.temp1}
\lefteqn{P_{g+h}(v_z|\phi,r,M_1,M_2) =}\nonumber \\ & &  
\int_{-1}^{\infty} P_{g+h}(v_z|\phi,r,M_1,M_2,\delta)
\,P_m(\delta|r,M_1,M_2)d\delta.
\end{eqnarray}

\noindent
To complete the calculation of the two-halo velocity PDF, equation
(\ref{e.temp1}) is integrated over all combinations of halo masses as
in equation (\ref{e.vpdf}), which gives $P_{2h}(v_z|\phi,r)$ (note
that the subscript {\it 2h} refers to two-halo galaxy pairs, as in \S
2). This PDF is used in the two-halo streaming model, equation
(\ref{e.stream2h}). Once the total velocity PDF has been calculated
and tabulated as a function of separation and angle, $\xi_{\rm
  1h}(r_\sigma,r_\pi)$ is calculated from equations (\ref{e.xi1h_cs})
and (\ref{e.xi1h_ss}) and $\xi_{\rm 2h}(r_\sigma,r_\pi)$ from equation
(\ref{e.stream2h}).


\begin{table*}
  \caption{Properties of the Simulations and HOD Parameters}
  \begin{tabular}{@{}ccccccc@{}}
  \hline
  {Model} & {$\s8$} & {$\mmin$ [\hmsol ]} & {$M_1$ [\hmsol ]} & {$\asat$} & {$\om(\beta=0.46)$} & {$\av$}\\
  \hline

HOD1 & 0.95 & 1.12$\times 10^{12}$ & 2.81$\times 10^{13}$ & 0.934 & 0.24 & 1.109\\
HOD2 & 0.90 & 1.12$\times 10^{11}$ & 2.75$\times 10^{13}$ & 0.959 & 0.26 & 1.076\\
HOD3 & 0.80 & 1.09$\times 10^{12}$ & 2.51$\times 10^{13}$ & 1.005 & 0.30 & 1.000\\
HOD4 & 0.70 & 1.08$\times 10^{12}$ & 2.24$\times 10^{13}$ & 1.109 & 0.36 & 0.906\\
HOD5 & 0.60 & 1.06$\times 10^{12}$ & 1.93$\times 10^{13}$ & 1.199 & 0.47 & 0.797\\
\hline
\end{tabular}
\medskip
\\ 
Note-- All masses have been scaled to $\om=0.3$. When we scale an
HOD to a different value of $\om$, the values of $\mmin$ and $M_1$
scale in proportion to $\om$. Values in the sixth and seventh columns
are the parameters of the HOD models used in \S 4.3, which have
$\beta=0.46$ and the same finger-of-god strength for all five models.

\end{table*}


\section{Results}

Although one-dimensional diagnostics are preferable when analyzing
observational measurements of redshift-space distortions, it is
worthwhile to demonstrate the ability of the model to reproduce the
overall features of the two-dimensional \xx. Figure \ref{xi2d_compare}
plots \xx\ for both the N-body results and model for our central
model, HOD3 in Table 1, plotted as contours of constant correlation in
the \spplane\ plane. The thick gray contours represent the N-body
data, averaged over the fifteen orthogonal projections of the five
realizations, and the thin black lines plot the model results. The
contours are separated by powers of 1.7, with the outermost contour
representing $1.7^{-5} = 0.07$. The coherent infall of matter into
overdense regions is present in the flattening of the contours on
large scales. In the absence of this infall, the contours would be
concentric circles at these scales. At small scales the FOG effect is
shown by the sharp elongation of contours at $r_\sigma \la 2$
\hmpc. The model accurately reproduces both the large- and small-scale
effects seen in the N-body results.

Although it is possible to compare model predictions directly to the
observed two-dimensional correlation function \xx, this is not
preferred due to the complexity of the error analysis. There is a
strong covariance between data points, making it difficult to
accurately estimate the covariance matrix of the measurements due to
the large number of data points. Using the two-dimensional data set
also makes it difficult to isolate constraints on the cosmological
parameters from constraints on the HOD parameters. We proceed instead
by extracting one-dimensional observables from the two-dimensional
data, observables that are more easily interpreted in terms of the
parameters $\om$, $\s8$, and $\av$.

\subsection{Redshift-space observables}

In Paper I we described three observables for the correlation
function. The first two, based on multipoles of \xx, emphasized
large-scale distortions because they have predictions in linear
theory. The multipoles of the correlation function are expressed as

\begin{equation}
\label{e.xi_l}
\xi_l(r) = \frac{2l+1}{2} \int_{-1}^{+1} \xi(r_\sigma,r_\pi) P_l(\mu) d\mu,
\end{equation}

\noindent 
where $r^2=r_\sigma^2+r_\pi^2$, $\mu=r_\pi/r$, and $P_l(\mu)$ is the
Legendre polynomial of order $l$. The ratio of the monopole to the
real-space correlation function,

\begin{equation}
\label{e.xizr}
\xi_{0/R}(r) = \frac{\xi_0(r)}{\xi_R(r)},
\end{equation}

\noindent
is predicted in linear theory to be a constant value larger than one,
but non-linearities in the velocity field suppress small-scale
redshift-space clustering, driving this ratio to zero as $r$ becomes
small. In Paper I, we restricted our use of this diagnostic to the
asymptotic value at $r\ge 10$ \hmpc, but small scales contain
additional information, and in this analysis we will make use of the
full range of $r$.

The quadrupole moment of the correlation function is defined as

\begin{equation}
\label{e.xiqp}
\xiQp(r) = \frac{\xi_2(r)}{\xi_0(r) - \bar{\xi}_0(r)}
\end{equation}

\noindent 
where $\bar{\xi}_0(r)$ is the volume averaged monopole,

\begin{equation}
\label{e.xi_bar}
\bar{\xi}_0(r) = \frac{3}{r^3}\int_0^r\xi_0(s) s^2 ds.
\end{equation}

\noindent
In the numerical results of Paper I, this diagnostic also has an
asymptotic value at large scales that is roughly consistent with its
linear theory prediction, but due to the higher-order multipole,
equation (\ref{e.xiqp}) is more sensitive to the galaxy velocity
dispersion, and non-linearities affect this diagnostic at larger $r$
than $\xizr$. For cosmologies that produce strong redshift
distortions, e.g. $\beta \ga 0.6$, $\xiQp$ does not reach a flat
plateau even at 40 \hmpc, putting the linear theory regime outside the
range of high precision measurements.

To quantify small-scale distortions, Paper I focused on the behavior of
$\xi(r_\pi)$ at fixed, small \rsig. We define the quantity $\rhalf$ as
the value of \rpi\ at which the correlation function falls by a factor
of two from its value at $r_\pi=0$, or

\begin{equation}
\label{e.rhalf}
\frac{\xi(\rhalf)}{\xi(0)} = \frac{1}{2}.
\end{equation}

\noindent
The quantity $\rhalf$ can be thought of as a characteristic length
scale for FOGs. At $r\sim 0.1$ \hmpc, $\rhalf$ depends on the
parameter combination $\om\av^2$ and is nearly independent of
$\s8$. At intermediate scales, $r \sim 1$ \hmpc, $\rhalf$ is highly
sensitive to $\s8$ because the number of one-halo pairs at this scale
depends on the shape of the halo mass function at large masses. For
the approach of Paper I, measuring $\rhalf$ at $r_\sigma=0.1$
\hmpc\ is beneficial because it provides a simple relation to
constrain $\om\av^2$.

The form of the HOD used in Paper I is simple in its parameterization;
halos above a minimum mass $\mmin$ have a central galaxy located at
the center of mass of the host halo, while satellite galaxies have a
mean occupation number that scales as a power law with the host halo
mass, i.e. $\nsat = (M/M_1)^\asat$. Therefore the mean occupation
function is

\begin{equation}
\label{e.hod}
\navg \equiv \ncen + \nsat = 1 + \left( \frac{M}{M_1} \right)^{\asat}
\end{equation}

\noindent
for $M\ge\mmin$, and $\navg=0$ below $\mmin$. The dispersion of
satellite galaxies about $\nsat$ is assumed to be Poisson, motivated
by both numerical results (\citealt{kravtsov04, zheng04b}) and
observational studies (\citealt{lin04}). The mean occupation of
central galaxies is a step function with no scatter. This
parameterization has only two free parameters\footnote{$\mmin$ is set
  by the space density of galaxies once $M_1$ and $\asat$ are chosen.}
but still has enough freedom to model the observed \wp\ of SDSS
galaxies (\citealt{zehavi04b}).  Table 1 lists the HOD parameters for
the five values of $\s8$ considered in Paper I, with all masses scaled
to $\om=0.3$. These parameters were chosen such that \xg\ was nearly
identical for all values of $\s8$. We make mock galaxy distributions
by populating the N-body halos with galaxies according to equation
(\ref{e.hod}). Satellite galaxies are placed randomly throughout the
halo following the appropriate NFW profile for each mass and
cosmology, calculated with the model of \cite{bullock01}. Velocities
are selected randomly from a Gaussian distribution with dispersion
given by equation (\ref{e.sigma_gal}). See Paper I for further
details.


\begin{figure}
\centerline{\psfig{figure=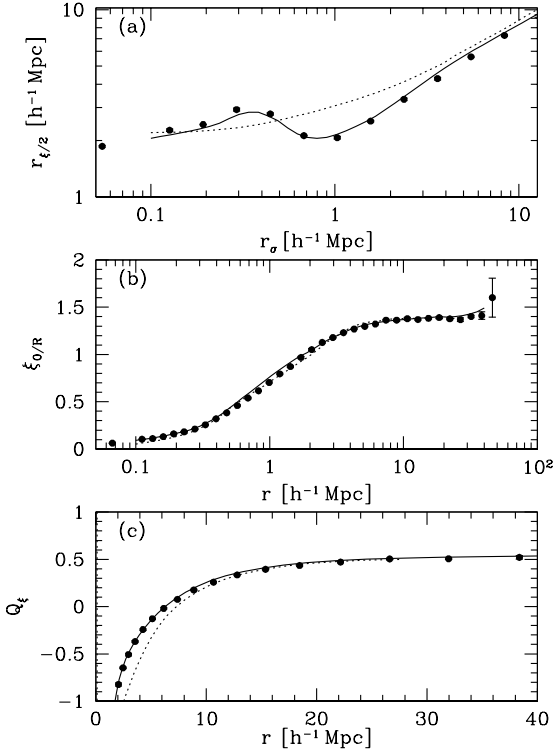,width=10.0cm}}
\caption{ \label{central_model}
One-dimensional measures of redshift-space clustering. In each panel, the
filled circles show the results of HOD3 from the N-body simulations (see
Table 1). Error bars are shown, but at most $r$ are smaller than the
point size. Solid lines show the model calculations, and dotted lines
show the best-fit dispersion model with $\sigma_v=418$
\kms. Panel (a) plots $\rhalf$ against \rsig. Panel (b) plots $\xizr$
against $r$. Panel (c) plots $\xiQp$ against $r$.}
\end{figure}

Figure \ref{central_model} plots the three redshift-space diagnostics
for the central model.  In all panels, the filled circles represent
the N-body data, the solid line plots the model calculation, and the
dotted line represents the dispersion model
(eq.~[\ref{e.lin_exp}])\footnote{We use the implementation of the
  dispersion model in configuration space as shown in
  \cite{hawkins03}.} for $\beta=0.46$ and the best-fit value of
$\sigma_v$ of 418 \kms. Note that our model does {\it not} have a
parameter (like $\sigma_k$) adjusted to match the redshift-space
distortions, and in that sense has less freedom to reproduce the data,
though the formulas describing the halo pairwise velocities were
calibrated on the simulations. Error bars on the N-body results are
calculated from the dispersion of the five simulations divided by
$\sqrt{4}$ to yield the error in the mean, but they are generally too
small to be seen outside the plot symbols.

Figure \ref{central_model}a plots the results for $\rhalf$ as a
function of transverse separation. Both the N-body results and the
model show a wave pattern that results from the transition in
\xg\ from the one-halo to two-halo term (see \S 5 in Paper I). As the
dominant source of pairs shifts from one-halo pairs in high-mass
systems to two-halo pairs in low-mass halos, the galaxy velocity
dispersion decreases sharply (\citealt{slosar06, tinker_pvd06}). The
dispersion model shows no such wave pattern and is a poor fit
to the N-body data, demonstrating the inability of a simple,
scale-independent velocity parameter to describe small-scale galaxy
motions.

Figure \ref{central_model}b plots $\xizr$ as a function of $\log
(r)$. The large-scale infall amplifies the clustering, making $\xizr>1$
at $r\ga 4$ \hmpc, while non-linearities suppress clustering, driving
$\xi_0/\xi_R$ to zero at small separation. Because $\xi_0(r)$ is the
lowest order multipole, it is least sensitive to the higher order
moments of the velocity field. Thus, both the HOD model and the
dispersion model provide good fits to the N-body results.

Figure \ref{central_model}c plots $\xiQp$ as a function of
$r$. Non-linearities become apparent in the quadrupole at a larger
scale than in the monopole; although $\xiQp$ flattens to a horizontal
asymptote at large scales, the data deviate significantly from this
asymptote at $r\la 20$ \hmpc, and $\xiQp$ becomes negative at $r\sim
6$ \hmpc. Both the HOD model and the dispersion model
reproduce the large-scale behavior, but the dispersion
solution is a poor fit at $r\la 10$ \hmpc. The HOD model is accurate
at all scales.


\begin{figure*}
\vspace{-2.0cm}
\centerline{\psfig{figure=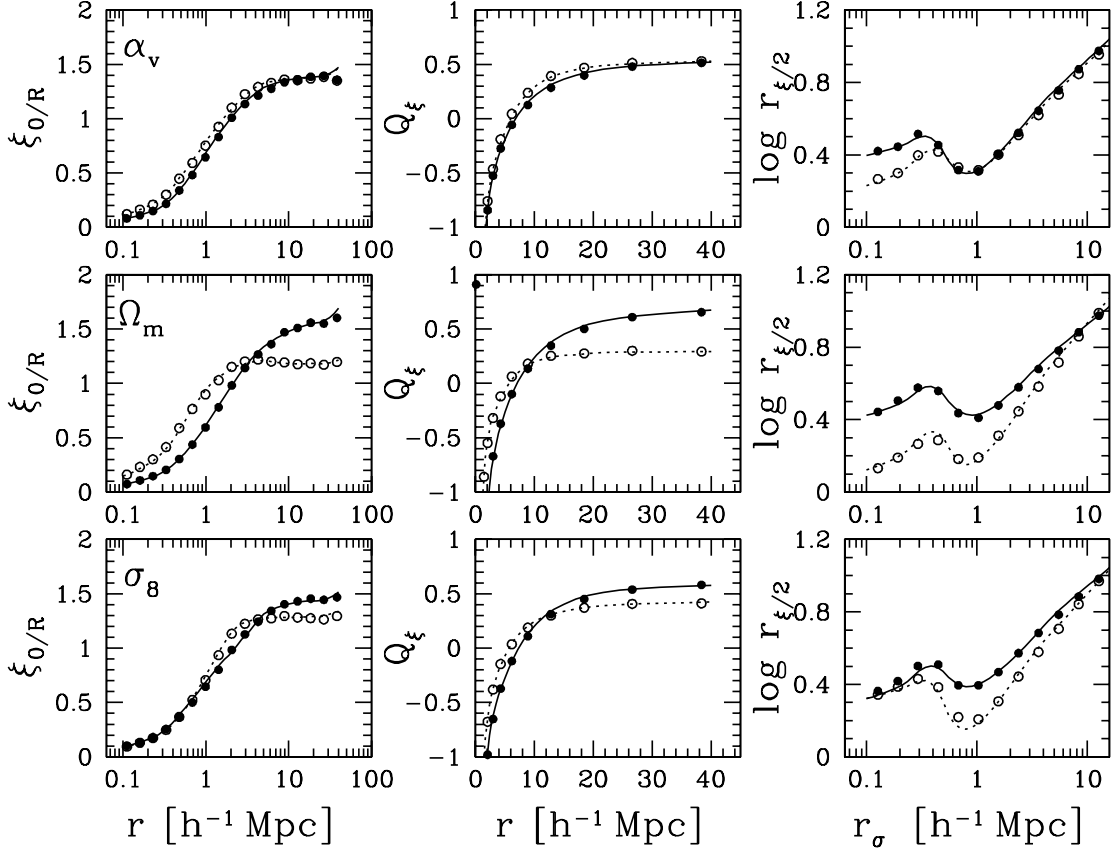,width=\two_col_fig}}
\caption{ \label{9win} Measures of redshift-space clustering for
  various cosmologies. In each panel, the circles represent the N-body
  results and the lines show the model results. In each row, a
  cosmological parameter is varied between the high and low values
  used in Paper I. The top row plots results for $\av=0.8$ (open
  circles; dotted lines), and $\av=1.2$ (filled circles; solid
  lines). The middle row plots results for $\om=0.1$ (open circles;
  dotted lines), and $\om=0.5$ (filled circles; solid lines). The
  bottom row plots results for $\s8=0.6$ (open circles; dotted lines),
  and $\s8=0.95$ (filled circles; solid lines). }
\end{figure*}

Figure \ref{9win} plots the results of the N-body simulations for
these same diagnostics but with variations in the $(\om, \s8, \av)$
parameter space. In each panel, the circles plot the N-body data, with
filled and open circles representing the high and low values of the
isolated parameter, respectively. The solid and dotted lines are the
model results for the high and low parameter values, respectively. The
bottom row shows the results for the highest and lowest values of
$\s8$ considered in Paper I (0.6 and 0.95), with $\om = 0.3$ and
$\av=1.0$. The middle row presents the extreme values of $\om$ (0.1
and 0.5), with $\s8=0.8$ and $\av=1$. The top row presents results for
setting $\av$ to 0.8 and 1.2 while keeping $\om$ and $\s8$ fixed at
their central values. In all cases, the HOD model accurately tracks
the changes in the redshift-space diagnostic with changes in
cosmological parameters, despite having no adjustable parameters of
its own.

\subsection{Parameter Recovery}

In Paper I we concluded that combining redshift-space information from
multiple scales can break degeneracies and allow determination of
$\om$, $\s8$, and $\av$. Models along the linear theory degeneracy
axis, $\om^{0.6}\s8={\rm const.}$, have nearly identical large-scale
redshift distortions that can mask significant changes in
cosmology. Models along the degeneracy axis controlling satellite
galaxy velocities, $\om\av^2={\rm const.}$, have the same value of
$\rhalf$ when measured at $r_\sigma \la 0.2$ \hmpc. Thus it is
possible for widely varying cosmologies to produce the same
large-scale anisotropies and the same small-scale FOGs. Despite these
degeneracies, information from intermediate scales, where the one-halo
to two-halo transition occurs, is able to break these degeneracies.


\begin{figure*}
\centerline{\psfig{figure=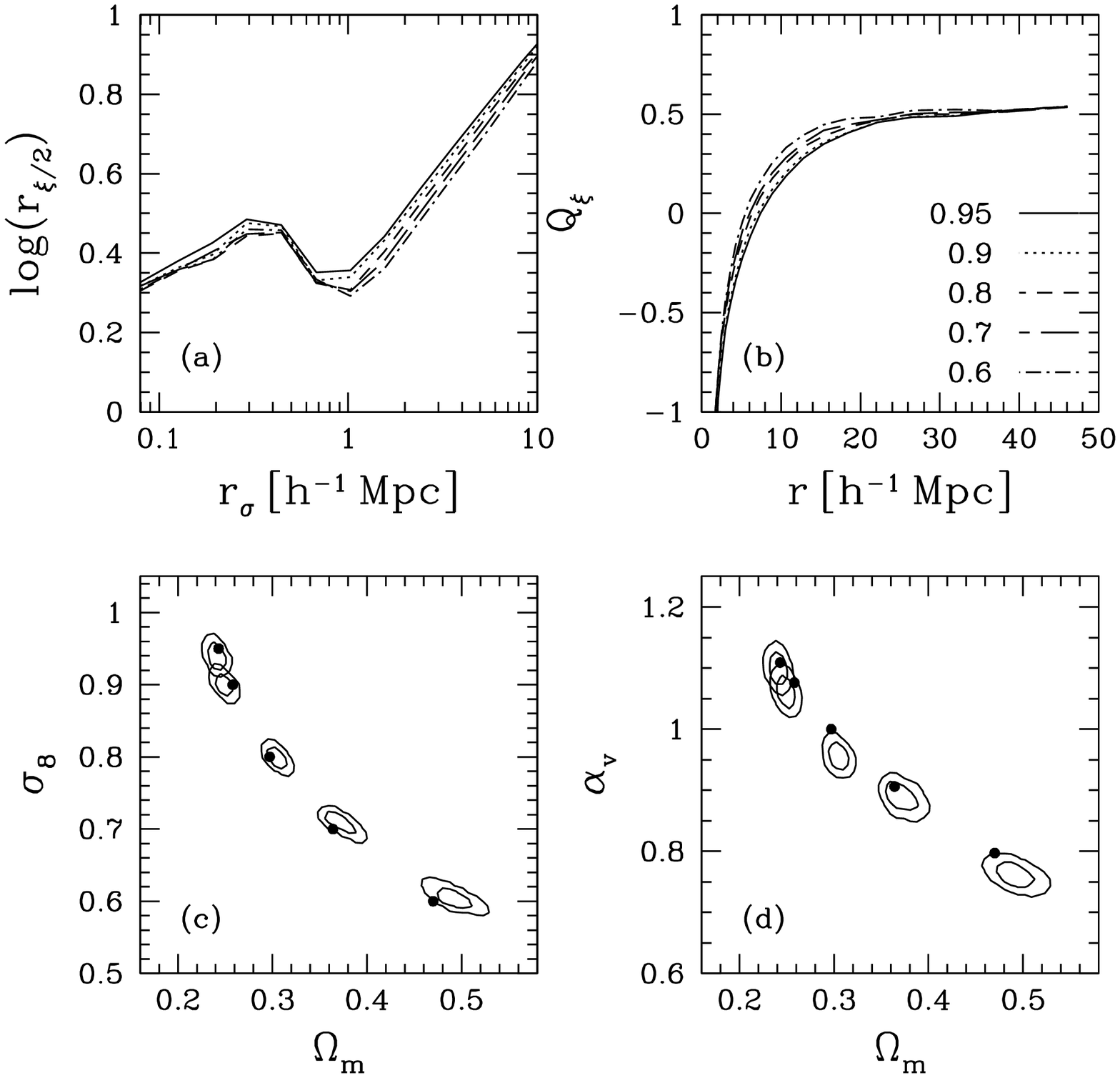,width=\two_col_fig}}
\caption{ \label{mcmc_results} Recovery of input cosmological
  parameters from MCMC analysis for the large-scale/small-scale
  degenerate HOD models with the parameter values listed in Table
  1. Panels (a) and (b) plot the input N-body data for two of the four
  observables used.  Panel (a) plots $\rhalf$ as a function of
  $r_\sigma$ for all five models. This quantity demonstrates the
  small-scale degeneracy of the models. Panel (b) plots $\xiQp$ as a
  function of $r$.  This quantity demonstrates the large-scale
  degeneracy of the models. Panel (c) plots the 1- and 2-$\sigma$
  error contours in the $\om$-$\s8$ plane, marginalizing over $\av$ and
  HOD parameters. The filled circles represent the input
  parameters. Panel (d) shows the error contours in the $\om$-$\av$
  plane, marginalizing over $\s8$ and HOD parameters.  }
\end{figure*}

To test the ability of the model to recover the correct cosmological
parameters, we use this set of large scale/small scale degenerate
models as the input data. The values of $\om$ and $\av$ for each
model are listed in columns 6 and 7 of Table 1. The models are scaled
to values of $\om$ such that $\beta=0.46$ for all five HODs, and
values of $\av$ are chosen such that $\rhalf \approx 2.1$ \hmpc\ at
$r_\sigma=0.1$ \hmpc.  All models have approximately the same
real-space correlation function (see Figure 2 in Paper I). We use four
observables; \xg , $\xizr$, $\xiQp$, and $\rhalf$, with diagonal error
bars calculated from the standard deviation of the mean of the five
simulations. We also add a systematic error to the model of 3\% in
each quantity to prevent data points with small errors from dominating
the $\chi^2$ minimization. With this systematic error, the $\chi^2$
per degree of freedom of the model calculation with the N-body data is
approximately one.

Figures \ref{mcmc_results}a and \ref{mcmc_results}b plot $\rhalf$ and
$\xiQp$, respectively, for the input N-body data. In panel (a), the
models produce degenerate $\rhalf$ values inside the one-halo term,
$r_\sigma< 0.7$ \hmpc. If $\av$ were set to unity for all five models,
the $\rhalf$ curves would separate at small radii due to the higher
velocity scale of satellite galaxies for the higher $\om$
models. Outside the one-halo term, the models separate modestly, but
are distinguishable outside their errors. In panel (b), all models
have the same asymptotic value of $\xiQp$, but all models are below
the linear theory prediction of 0.55 at all $r$ for $\beta =
0.46$. The different values of $\av$ for each model cause the radii at
which non-linearities set in to differ; for models with lower values of
$\av$, $\xiQp$ becomes negative at smaller $r$.


\begin{figure*}
\centerline{\psfig{figure=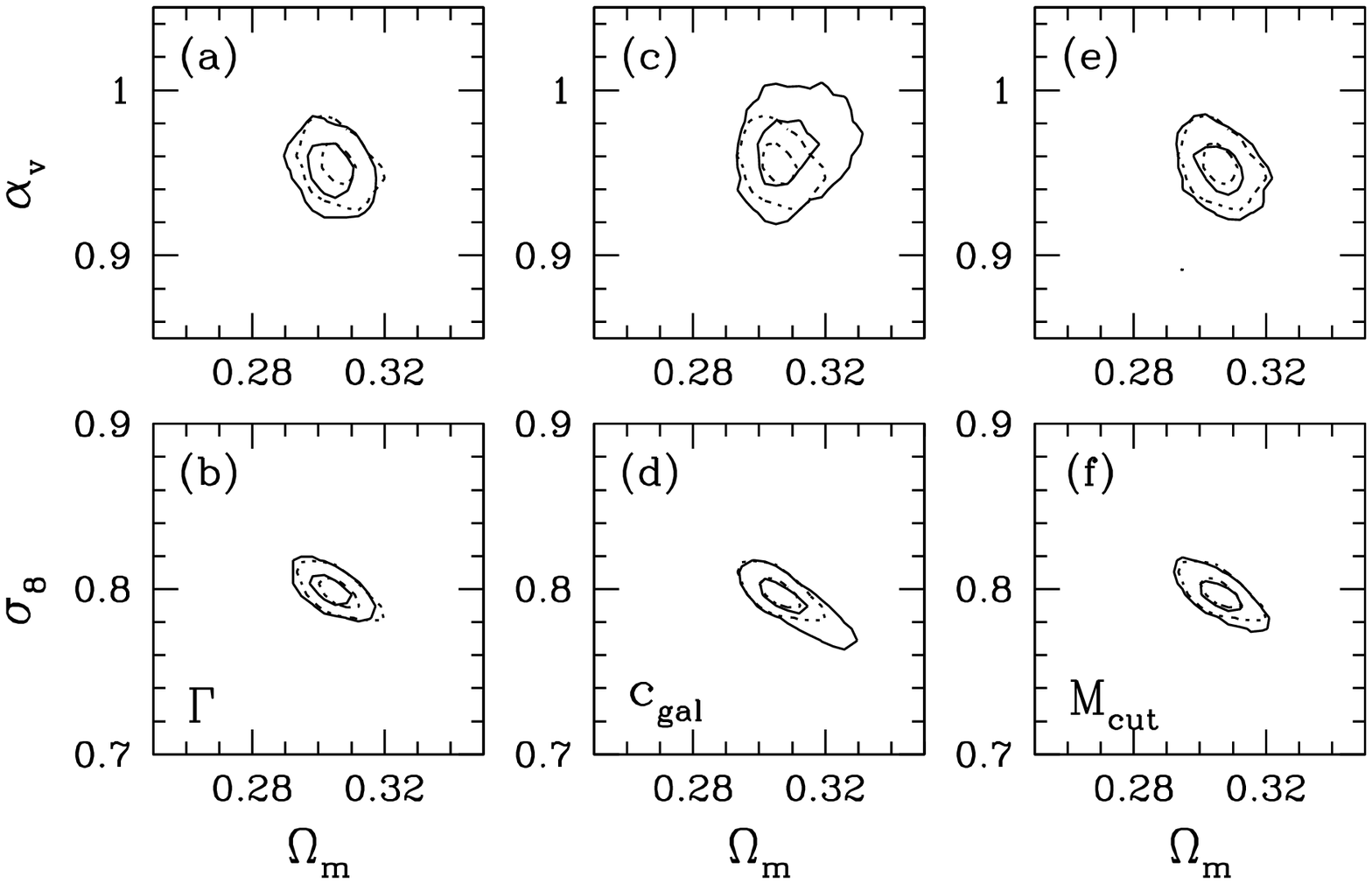,width=15.0cm}}
\vspace{-3.7cm}
\caption{ \label{extra_params} Constraints on $\om$, $\s8$, and $\av$
  for models with extra free parameters. In each panel, the dotted
  contours represent the error contours from Figure \ref{mcmc_results}
  for HOD3. The top three panels show constraints in the $\om$-$\av$
  plane, while the bottom panels show constraints in the $\om$-$\s8$
  plane. In panels (a) and (b), the solid lines correspond to a model
  in which $\Gamma$ is introduced as a free parameter, in addition to
  the HOD and cosmological parameters. In panels (c) and (d), the
  solid lines correspond to a model in which the ratio between the
  halo concentrations and the concentration parameter for satellite
  galaxies, $c_{\rm gal}/c_{\rm halo}$ is a free parameter. In panels
  (e) and (f), the solid lines correspond to a model in which $\nsat$
  contains an extra parameter $\mcut$ (see equation
  [\ref{e.nsat_mcut}]).}
\end{figure*}

To explore the likelihood of cosmological parameters from these data
we use the Monte Carlo Markov chain (MCMC) technique (e.g.,
\citealt{doran04}). In MCMC, a randomly chosen point in parameter
space is automatically accepted as part of the chain if the $\chi^2$
for that point is less than that of the previous point. If the
$\chi^2$ is larger, the probability of acceptance is $\exp(-\Delta
\chi^2/2)$. To choose the next point in parameter space, parameters
are chosen from a Gaussian distribution centered on the previously
accepted point. The widths of the Gaussians are the eigenvalues of the
covariance matrix of the points already in the chain. Each chain is
checked for convergence using the power spectrum method of
\cite{dunkley05}. The benefit of the MCMC approach is that it allows
one to marginalize over the HOD parameters; because different HODs
will produce different galaxy velocity fields, it is necessary to
account for uncertainties in the HOD parameters from the errors in the
real-space correlation function. For each element in the chain, the
total $\chi^2$ is the sum of the $\chi^2$ values for each of the four
observables.

For this analysis we have altered the parameters of the Jenkins mass
function to better match the mass function of our own simulations. Due
to shot noise in the particle distribution, the friends-of-friends
algorithm systematically overestimates the masses of halos with $N$ less
than a few hundred particles (\citealt{warren05}). At fixed (true) mass,
this leads to an overestimate of the mass function by a few
percent. As shown in Appendix B of \cite{tinker05a}, the slight
difference between our mass function and the Jenkins fitting function
creates a discrepancy between the numerical and analytic one-halo
terms of \xg. In the MCMC analysis, this discrepancy can drive the
chain to the wrong solution even though the difference in the model
results is small. We stress that this modification is not a
consequence of the model; the model will correctly describe the
redshift-space distortions for any input mass function. The parameters
of the Jenkins formula are $(0.315, 0.61, 3.8)$. Our modified
parameters are $(0.345, 0.636, 3.43)$, which increases the number of
halos less than \mstar\ by $\sim 7\%$. The use of simulations with
higher mass resolution would produce mass functions in better
agreement with the results of \cite{jenkins01} and \cite{warren05}.

Figures \ref{mcmc_results}c and \ref{mcmc_results}d show the
constraints on the cosmological parameters from the MCMC results.
Panel (c) shows the $\om$-$\s8$ projection of the parameter space, and
panel (b) panel rotates the data to show the $\om$-$\av$
projection. The solid circles plot the input cosmological values. The
contours show the one- and two-$\sigma$ confidence levels on each
parameter. For most of the models listed in Table 1, the correct
values are within either the one- or two-$\sigma$ error contour. For
the central model, the MCMC analysis correctly yields $\om=0.306 \pm
0.006$ and $\s8 = 0.799 \pm 0.009$, but the velocity bias $\av = 0.955
\pm 0.015$ is 3-$\sigma$ off the true value of 1.0. For models HOD1,
HOD2, and HOD4, the analytic model correctly recovers the true $\av$
to within 1-$\sigma$ ($\sim 1.5\%$). The source of this discrepency in
HOD3 can be traced to \xg; the analytic model underpredicts the
amplitude of \xg\ for this model at $r\approx 1$ \hmpc\ (see Appendix
B in \citealt{tinker05a} for a comparison between the analytic model
and N-body results for this HOD). To produce a better fit to the
real-space clustering, a higher value of $\asat$ is favored ($1.032$
versus $1.005$). This increases the mean velocity dispersion of
galaxies, so the likelihood is higher for a model with slightly
negative velocity bias, $\sim 4.5\%$ below the true value. This
discrepancy is not significant in model the other models (we will
discuss the results for HOD5 presently). In tests with N-body HOD
models tailored toward samples of brighter, less abundant galaxies,
this discrepancy is absent in the analytic model for \xg.

Compared to HOD models 1 through 4, which recover the input cosmology
to better than $\sim 2\%$ (with the exception discussed in the
previous paragraph), the recovered parameters for HOD5 are barely
within the 2-$\sigma$ contours. The analysis yields $\Omega_m =
0.491\pm 0.014$, $\s8 = 0.607 \pm 0.010$, and $\av = 0.765 \pm 0.012$,
compared with the input values of $(0.47, 0.6, 0.797)$. The high
preferred value of $\om$ drives the preferred value of $\av$ lower
than the input value. The most likely value of $\s8$, however, is
unaffected.  The larger errors for this model could represent a
breakdown in $\om^{0.6}$ as an approximation to the linear theory
scaling of halo velocities or the inability of linear theory itself to
accurately model the scaling. Further testing is required to isolate
the source of the error, although it is likely outside the range of
interest for the true value of $\om$.

\subsection{Adding freedom to the model}

The test in the previous section made several simplifying
assumptions. First, the shape of $\plin$ was fixed to the true value
from the simulations. Second, the radial profiles of the satellite
galaxies were assumed to follow the true profiles used to populate the
simulations. Third, the functional form of the HOD used in the
analysis was the same as that used to create the test data.

In Figure \ref{extra_params}, we test the robustness of the analytic
model against these assumptions. Panels (a) and (b) compare the
results of the MCMC analysis from Figure \ref{mcmc_results} to an
analysis in which the power spectrum shape parameter $\Gamma$ is added
as an additional free parameter with no prior. The dotted contours
represent the 1- and 2-$\sigma$ constraints on the cosmological
parameters from Figure \ref{mcmc_results}, in which $\Gamma$ is set to
0.2. The solid contours are the results once marginalized over
$\Gamma$. The constraints are nearly identical, even with the added
freedom to the model, and the most likely value of $\Gamma = 0.204 \pm
0.005$. As discussed in \S 1, altering the shape of $\plin$ has
limited impact on the redshift-space observables employed here. But
altering $\Gamma$ changes the amplitude of \xg\ for a fixed HOD, so
large deviations from the true value of $\Gamma$ are easily excluded
by the real-space clustering alone.

Figures \ref{extra_params}c and \ref{extra_params}d show the
cosmological constraints of the analytic model when the galaxy density
profile within halos is allowed to vary. In this model, we introduce a
new parameter $a_c=c_{\rm gal}/c_{\rm halo}$, where $c$ represents the
concentration parameter of the NFW density profile, and $c_{\rm halo}$
is taken from \cite{bullock01}. This parameter preserves the shape of
the dependence of $c$ on halo mass, but allows its normalization to
vary. Panel (c) presents the constraints in the $\om$-$\av$ plane from
Figure \ref{mcmc_results} and from varying-$a_c$ model. Marginalizing
over $a_c$ has a noticeable effect on the constraints. The 1-$\sigma$
errors on $\om$ increase from $0.006$ to $0.009$, while the 1-$\sigma$
errors on $\av$ increase from $0.015$ to $0.021$. Although the most
likely values of $\om$ and $\av$ are consistent in the solid and
dotted contours, marginalizing over $a_c$ does not expand the contours
in a symmetric fashion. As discussed in the previous section, the
analytic model for HOD3 underpredicts \xg\ at $r=1$ \hmpc. When
marginalizing over $a_c$, this discrepancy with \xg\ can be
compensated for by decreasing the concentrations of the satellite
galaxies, thereby shifting one-halo pairs out to larger
separation. Thus, in the varying-$a_c$ analysis, the preferred value
of $\asat$ is closer to the true value of 1.005, skewing the error
contours, but the constraints on $\asat$ are somewhat weaker than in
the fixed-$a_c$ analysis, yielding larger overall errors. The
1-$\sigma$ constraints on $a_c$ are $0.93 \pm 0.16$. Panel (d)
presents the constraints in the $\om$-$\s8$ plane for these two
models. The 1-$\sigma$ error contours are the same in the two models,
but the extra freedom of the varying-$a_c$ model produces a small
extension of the 2-$\sigma$ contour down the $\om^{0.6}\s8$ degeneracy
axis. The weaker constraints produced by the varying-$a_c$ model are
apparent in Figure \ref{extra_params}, but they only increase the
errors from $\sim 2\%$ to $\sim 3\%$ on the cosmological parameters.

Lastly, we demonstrate the robustness of the model to extra freedom in
the parameterization of the HOD itself. The satellite occupation
function used throughout this paper is a simple power law with two
free parameters. In Figures \ref{extra_params}e and \ref{extra_params}f
we compare the MCMC results from our two-parameter $\nsat$ to a model
with a three-parameter $\nsat$ of the form

\begin{equation}
\label{e.nsat_mcut}
\nsat = \left( \frac{M - \mcut}{M_1} \right)^\asat,
\end{equation}

\noindent
where $\mcut$ is a cutoff mass scale. In equation (\ref{e.hod}),
$\nsat$ is cut off sharply at $\mmin$. The form of equation
(\ref{e.nsat_mcut}) has a soft cutoff that can be at any mass,
although halos below $\mmin$ are not allowed to host satellites. (Even
with this restriction, models with $\mcut<\mmin$ are not degenerate
until $\mcut\ll \mmin$.)  However, this extra freedom in the
occupation function produces a negligible effect on the cosmological
constraints. Panels (e) and (f) once again compare the error contours
in $\om$-$\av$ and $\om$-$\s8$ planes, respectively, for the standard
two-parameter $\nsat$ and varying-$\mcut$ model. The contours are
nearly unchanged between the two models.


\begin{figure}
\centerline{\psfig{figure=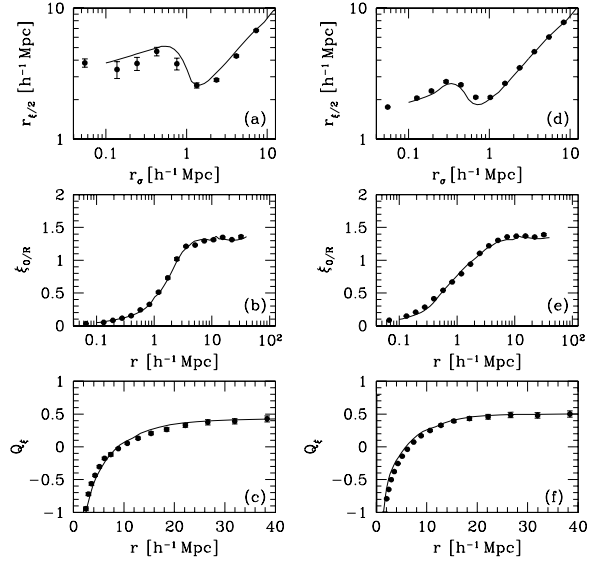,width=\one_col_fig}}
\caption{ \label{gamma_hod} Left-hand panels plot the standard
  redshift-space observables for HOD parameters constrained to match
  the $\mrlogh<-21$ SDSS \wp. The filled circles are N-body results
  for $\rhalf$ in panel (a), $\xizr$ in panel (b), and $\xiQp$ in
  panel (c). The solid lines are the analytic model calculations for
  each quantity for the same HOD parameters. Right-hand panels plot
  the same observables, but for a model in which $\Gamma=0.12$. The
  cosmology and HOD parameters are the same as HOD3 in Table 1. Points
  and lines are the same as panel (a)--(c). }
\end{figure}

\subsection{Other mock samples}

The HOD parameters in Table 1 produce galaxy space densities
consistent with the $\mrlogh<-20$ SDSS sample
(\citealt{zehavi04b}). Although our N-body simulations lack the
resolution to model fainter samples, we can test the accuracy of the
model against mock data matching to brighter samples. In Figures
\ref{gamma_hod}a-\ref{gamma_hod}c, we compare the analytic model to
N-body data created with HOD parameters determined from fitting
\wp\ measurements of the $\mrlogh<-21$ sample from
\cite{zehavi04b}. To populate the simulations, we use the HOD
parameters listed in Table 3 of \cite{zehavi04b}: $\mmin = 5.25\times
10^{12}$ \hmsol, $M_1 = 12.3\times 10^{14}$ \hmsol, and $\asat =
1.39$. These HOD parameters are used to populate the simulations with
$\s8=0.9$, and the halo velocities are scaled to $\om=0.3$. This
sample yields a space density lower than the HODs in Table 1 by a
factor of $\sim 5$, and emphasizes higher halo masses. The N-body
results are shown with the filled circles for $\rhalf$ in panel (a),
for $\xizr$ in panel (b), and $\xiQp$ in panel (c). The model
calculations for the same cosmology and HOD are shown with the solid
lines in each panel. For $\rhalf$, the lower space density of galaxies
increases the error on this diagnostic within the one-halo term,
partially from Poisson noise and partially from a less robust
determination of the asymptotic value of $\xi(r_\pi)$ at small
\rpi. But the analytic model compares favorably to the data, correctly
predicting the location and depth of the minimum at $r_\sigma = 1.5$
\hmpc. The analytic model also accurately predicts the shape of
$\xizr$, which has a sharper transition from linear to non-linear
scales than the results in Figure \ref{central_model}b due to the
emphasis on higher mass halos. For $\xiQp$, the analytic model
correctly describes the large-scale asymptote as well as the scale at
which non-linearities set in.

In Figures \ref{gamma_hod}d--\ref{gamma_hod}f, we compare the analytic
model to N-body data created with the HOD parameters and cosmology of
the central model, but populating the halos of simulations in which
the shape parameter of the linear matter power spectrum is
$\Gamma=0.12$.\footnote{The specifics of these simulations are
  identical to the $\Gamma=0.2$ simulations. See Paper I for more
  discussion.} Although a power spectrum of this shape is unable to
yield acceptable HOD fits to current SDSS measurements of
\wp\ (Z. Zheng, private communication), it is worth testing the
analytic model against an extreme change in $\plin$. Changing the
shape of $\plin$ alters not only the halo mass function and bias
relation but the one-point distribution of dark matter densities used
in the velocity model as well. For all three observables, the analytic
model performs well in modeling the redshift-space clustering for a
substantially different power spectrum.

This success is a strong indication that the analytic model
is correctly describing the underlying physics of redshift-space
distortions as it accurately describes a cosmological model
quite different from the one on which it was calibrated.


\section{Summary}

We have presented a model for the redshift-space correlation function
of galaxies using the framework of the halo occupation distribution
(HOD). The HOD quantifies galaxy two-point clustering by separating
the total number of galaxy pairs into the contributions from within a
single halo and pairs from two distinct halos. In redshift space, the
one-halo term is obtained by assuming galaxies follow an isothermal,
Gaussian velocity distribution with a dispersion proportional to the
virial dispersion of the dark matter halo. The constant of
proportionality $\av$ represents the velocity bias of galaxies, and it
encompasses the effects of non-isothermality, anisotropy, or systems
being unrelaxed. Central galaxies are assumed to follow the center of
mass of the halo they occupy. Pairs of galaxies from two distinct
halos move with a combination of virial motions and the halo center of
mass velocity. Halo velocity statistics have significant third and
fourth moments which must be modeled to calculate redshift-space
clustering. The key simplification that makes our analytic model
tractable is that the pairwise distribution of halo velocities at
fixed density is approximately Gaussian and approximately independent
of halo mass.  The non-Gaussian shape of the full distribution arises
from averaging over pairs in different density environments, and the
dependence of the distribution on halo mass arises because higher mass
halos tend to reside in denser environments.

Our approach solves several complications in modeling pairwise
velocity statistics at non-linear and quasi-linear scales. It
demonstrates why halo velocity PDFs in simulations are distinctly
non-Gaussian, with significant skewness and kurtosis that fluctuate
with scale. The skewness arises from halo pairs in high-density
environments, where pairs have high relative velocities (see also
\citealt{juszkiewicz98}). The kurtosis results from a superposition of
Gaussians of varying widths from halo pairs in wide-ranging
environments. The model also circumvents to problem of correlated
radial and tangential velocities. At a given $\delta$, $v_r$ and $v_t$
are independent, and it is straightforward to calculate $v_z$ at any
angle with respect to the observer's line of sight. The model is
highly flexible in that it can describe redshift-space clustering for
samples of any space density without the associated complications
arising from N-body simulations that sacrifice volume for mass
resolution, or resolution in favor of box size. It also eliminates the
need to run large numbers of simulations when exploring parameter
space.

This model can be used to obtain constraints in the $(\om, \s8, \av)$
parameter space from redshift-space anisotropies following the
blueprint detailed in Paper I. Observations of \wp\ provide tight
constraints on the parameters of the HOD, separating the constraints
on bias from the constraints on cosmology. Our choices for
redshift-space observables largely normalize out the shapes of the
real-space galaxy correlation function and the linear matter power
spectrum. Marginalization over $\plin$ and HOD parameters do not
degrade the ability of the model to constrain parameters. The only
unknown quantity that notably affects these constraints is the profile
of satellite galaxies within groups $c_{\rm gal}$, a quantity that can
be observationally measured (\citealt{yang05}) and theoretically
predicted (e.g., \citealt{diemand04, nagai05}). Even with no prior on
$c_{\rm gal}$, cosmological constraints only increase by $\sim 1\%$
relative to the true value.

Our HOD approach represents a significant improvement over the
dispersion model (Eq.~[\ref{e.lin_exp}]) for two main reasons. First,
the HOD model circumvents the use of the redshift-space distortion
parameter $\beta$ entirely. The cosmological parameters $\om$ and
$\s8$ are direct inputs of the model, eliminating the degeneracy
involved with using $\beta$ to determine the matter density. Second,
there are no free parameters like $\sigma_k$, which creates a
degeneracy with $\beta$ as well. The velocity bias parameter $\av$ is
a truly physical parameter that is cosmologically relevant and is a
testable prediction of numerical simulations of galaxy formation and
semi-analytic models. Measurements of $\av$ can reduce uncertainty in
dynamical mass estimates of galaxy groups, utilized in group catalogs
culled from large-scale optical surveys (e.g., the 2PIGG catalog of
\citealt{eke04}). Alternately, studies that use the abundance of
systems as a function of their velocity dispersion require knowledge
of $\av$ to provide useful comparisons to theoretical models
(\citealt{davis04, gerke05}). In this paper we have assumed that
central galaxies are not subject to virial motions. Modest values of
$\avc$ have limited effect on redshift-space clustering, but non-zero
$\avc$ can easily be incorporated into the analysis of observational
data and either marginalized over or constrained if the data are of
sufficient precision.

Utilizing clustering information on multiple scales breaks the two
primary degeneracy axes in redshift-space distortions. At large
scales, linear theory correctly predicts that models with the same
value of $\beta$ will have the same large-scale distortions (even
though linear theory is a poor descriptor of the shape of the
distortions for a given value of $\beta$). This creates the degeneracy
$\om^{0.6}\s8={\rm const}$. At small scales, non-linearities are
determined by the velocity scale of dark matter halos, creating the
degeneracy axis $\om\av^2={\rm const}$. Paper I concluded that
intermediate-scale clustering, $1$ \hmpc\ $\la r \la$ 10 \hmpc,
provides the key for breaking the large-scale/small-scale
degeneracies. Here we have quantitatively demonstrated this
result. The typical ``banana-curve'' degeneracy axes can be seen in
Figures \ref{mcmc_results}c and \ref{mcmc_results}d, but the
simulations are clearly distinguished from one another by the analytic
model. The large-scale distortions set the amplitude of the
$\om^{0.6}\s8$ degeneracy curve, the small-scale distortions normalize
the $\om\av^2$ degeneracy curve, and the intermediate-scale clustering
locates the model along both curves.

The current data releases of the SDSS and 2dFGRS have reached the
necessary volume and statistics to implement this HOD analysis. The
ability to extract information from galaxy clustering measurements at
quasi-linear and non-linear scales enables precise constraints on our
specified subset of cosmological parameters that can be competitive
with large-scale measures such as CMB anisotropies, Lyman-$\alpha$
forest, and the large-scale galaxy power spectrum. Although
constructed on more fundamental physics, these methods are sensitive
to quantities like the CMB tensor-to-scalar ratio, mass of the
neutrino, and curvature of the inflationary fluctuation spectrum. For
small-scale galaxy clustering, modest changes in the shape of $\plin$
have marginal impact on halo occupation constraints. Thus
redshift-space clustering measurements provide a powerful and
complementary method for estimating the amplitude of dark matter
clustering and the mean matter density of the universe.

\vspace{1cm}
\noindent I would like to thank Andreas Berlind, Juna Kollmeier,
Andrey Kravtsov, Roman Scoccimarro, Jaiyul Yoo, Andrew Zentner, and
Zheng Zheng for useful discussions, and Volker Springel for providing
the public \gadget\ code. I would especially like to thank David
Weinberg for his help with many aspects of this paper. The simulations
were performed on the Beowulf and Itanium clusters at the Ohio
Supercomputing Center under grants PAS0825 and PAS0023. I acknowledge
the support of a Distinguished University Fellowship at the Ohio State
University during the course of this work. This work was also
supported by NSF grants AST-0098584 and AST-0407125.


\appendix
\section{Outline of Two-Halo Term}

For the convenience of readers who wish to implement our model, we
present a concise outline of the calculation of $\xi_{\rm
  2h}(r_\sigma,r_\pi)$, beginning with the definition of the streaming
model,

\begin{equation}
\label{ea.stream2h}
1+\xi_{\rm 2h}(r_\sigma,r_\pi) = \int_{-\infty}^\infty [1+\xid(r)]\, P_{\rm 2h}(v_z|r,\phi)\, dv_z,
\end{equation}

\noindent 
where $r^2=r_\sigma^2+z^2$, $v_z =H\,(r_\pi-z)$, and $\cos \phi =
r_\sigma/r$. $\xid(r)$ is the two-halo contribution to the real-space
correlation function, and $P_{\rm 2h}(v_z|r,\phi)$ is the
line-of-sight velocity PDF of galaxy pairs from two distinct halos. The
velocity PDF in equation (\ref{ea.stream2h}) is a pair-weighted average
over all possible combinations of halo masses. For a given halo pair,
$M_1$ and $M_2$, the line-of-sight PDF is an integral over all possible
densities in which that pair can exist.

The line-of-sight PDF of a halo pair at density $\delta$ is

\begin{eqnarray}
\label{ea.pdf_sum}
\lefteqn{P_{g+h}(v_z|\phi,r,M_1,M_2,\delta) = } \nonumber \\ & & 
\sum_{i=1}^4\, \frac{w_i}{\sqrt{2\pi}\sigma_i\,} \exp\left[ \frac{-(v_z - \mu_r\sin\phi)^2}{2\sigma_i^2}
\right],
\end{eqnarray}

\noindent
where each term in the summation represents a different combination of
central and satellite galaxy pairs from the two halos. The dispersion for each term is

\begin{equation}
\label{ea.dispersion}
\sigma_i^2 = (\sigma_t^2\cos^2\phi + \sigma_r^2\sin^2\phi) 
	+ [\sigma_{\rm gal}^{(i)}]^2,
\end{equation}

\noindent
where $\sigma_t$ and $\sigma_r$ are the tangential and radial
dispersions of the dark matter halos, and $\sigma_{\rm gal}^{(i)}$ is
the dispersion of the galaxy pair from internal halo motions, defined
by equation (\ref{e.sigma_gal}). The radial dispersion is

\begin{equation}
\label{ea.sigma_v}
\sigma_{r} = 200 \left(\frac{\om}{0.3}\right)^{0.6}\left(\frac{\s8}{0.8}\right)
\left(\frac{\rhob}{\rhob_{200,r}}\right)^\alpha {\rm km\,s^{-1}},
\end{equation}

\noindent where 

\begin{eqnarray}
\label{ea.arho200r}
\lefteqn{\rhob_{200,r}(M_1,M_2,r) =}\nonumber \\ & &  
\left(\frac{r}{5.0\rvira^{1/2}}\right)^{-4.0} + 
\left(\frac{r}{11.5\rvirc^{1/2}}\right)^{-1.3} + \,\,0.50,
\end{eqnarray}

\noindent and 

\begin{equation}
\label{ea.alpha}
\alpha(r) = \left(\frac{r}{35 h^{-1}{\rm Mpc}}\right)^{0.1}.
\end{equation}

\noindent 
For the tangential velocities, $\rhotwor$ is replaced in equation
(\ref{ea.sigma_v}) with

\begin{eqnarray}
\label{ea.arho200t}
\lefteqn{\rhob_{200,t}(M_1,M_2,r) =}\nonumber \\ & & 
\left(\frac{r}{7.2\rvira^{1/2}}\right)^{-2.5} + 
\left(\frac{r}{12.6\rvirc^{1/2}}\right)^{-0.8} + \,\,0.48.
\end{eqnarray}

\noindent
The units of $r$ and $\rvir$ are \hmpc. The mean velocity $\mu$ of
equation (\ref{ea.pdf_sum}) is given by the spherical collapse model
(eq.~[\ref{e.mu_sc}]) at $r\le 4$ \hmpc, linear theory
(eq.~[\ref{e.mu_linear}]) at $r\ge 20$ \hmpc, and a linear combination
of the two at intermediate scales,

\begin{equation}
\label{ea.mu}
\mu(r,\delta) = w\,\mu_{\rm sc} + (1-w)\,\mu_{\rm lin},
\end{equation}

\noindent with

\begin{equation}
\label{ea.mu_weight}
w = \left\{ \begin{array}{ll}
		1 & {\rm if\ \ } r \le 4 \\
		1.86 - 0.62\,\ln\,r & {\rm if\ \ } 4 \le r < 20 \\
		0 & {\rm if\ \ } r < 20 \,\,.\\
		\end{array}
	\right.
\end{equation}

The distribution function $\Pgh(v_z|r,\phi,M_1,M_2,\delta)$ is
integrated over all $\delta$ to obtain

\begin{eqnarray}
\label{ea.int_over_delta}
\lefteqn{P_{g+h}(v_z|\phi,r,M_1,M_2) =} \nonumber \\ & &  
\int_{-1}^{\infty}\, P_{g+h}(v_z|\phi,r,M_1,M_2,\delta)
P_m(\delta|r,M_1,M_2)d\delta,
\end{eqnarray}

\noindent 
where the PDF of dark matter density used in equation
(\ref{ea.int_over_delta}) is given by

\begin{eqnarray}
\label{eq.density_pdf}
\lefteqn{P_m(\delta|r,M_1,M_2) = } \nonumber \\ & & A\exp\left[\frac{1.45(b_1+b_2) +
(r/9.4\rvira)^{-2.2}} {1+\delta}\right]P_m(\delta|r) \nonumber \\
\end{eqnarray}

\noindent and

\begin{equation}
\label{ea.lognormal}
P_m(\delta|r) = \frac{1}{2\pi\sigma_1^2} \, \exp \left[- 
\frac{[\ln(1+\delta) + \sigma_1^2/2]^2}{2\sigma_1^2} \right] \frac{1}{1+\delta},
\end{equation}

\noindent where $\sigma_1^2(r) = \ln [1+\sigma_m^2(r)]$ and
$\sigma_m(r)$ is the mass variance in top-hat spheres of radius
$r$. To calculate $\sigma_m(r)$, we use the non-linear matter power
spectrum of \cite{smith03} to better describe the distribution of dark
matter at small scales. To calculate the halo bias factors, we use the
function in Appendix A of \cite{tinker05a}. The final step is to
integrate over all halo pairs,

\begin{eqnarray}
\label{ea.vpdf}
\lefteqn{P_{\rm 2h}(v_z|r,\phi) = 2(\ngavgp)^{-2} \int_0^{\mlima} dM_1
\frac{dn}{dM_1} \navga} \nonumber \\ & &
\int_0^{\mlimb} dM_2 \frac{dn}{dM_2} \navgb
P_{\rm g+h}(v_z|r,\phi,M_1,M_2) \,,
\end{eqnarray}

\noindent 
where $\ngavgp$ is given by equation (\ref{e.ng_sph}), and the limits of
the integral are determined by the condition that no halo pair can be
closer than the sum of their virial radii. Using the relation $\rvir =
(3M/4\pi\bar{\rho}\Delta_{\rm vir})^{1/3}$ (recall that we have defined
$\Delta_{\rm vir}=200$ throughout this paper),

\begin{equation}
\begin{array}{lll}
\mlima(r) & = & r - \rvir(\mmin) \\
\mlimb(r) & = & r - \rvir(M_1) \, . \\
\end{array}
\end{equation}

\noindent 
Equation (\ref{ea.vpdf}) is inserted directly into equation
(\ref{ea.stream2h}).



\end{document}